%% file: main.tex
\documentclass{article}
\usepackage[a4paper]{geometry}
\usepackage[round]{natbib}
\usepackage[dvipsnames]{xcolor}
\usepackage[colorlinks,linktoc=true,
             citecolor= Cerulean!85!green!90!black,
             linkcolor=YellowOrange,
             urlcolor=RubineRed!85!black,
             ]{hyperref}

\usepackage{amssymb}
\usepackage{amsmath}
\usepackage{amsthm}

\usepackage{graphicx}
\usepackage{subfig}
\usepackage{multirow}
\usepackage{enumitem}

\usepackage{authblk}

\usepackage{soul}

\usepackage{cleveref}
\crefformat{section}{\S#2#1#3} 
\crefformat{subsection}{\S#2#1#3}
\crefformat{subsubsection}{\S#2#1#3}

\newtheorem{theorem}{Theorem}[section]
\newtheorem{lemma}[theorem]{Lemma}
\newtheorem{prop}[theorem]{Proposition}

\DeclareMathOperator{\Var}{\widehat{Var}}

\newcommand{\wh}{\hat}

\newcommand{\R}{\mathbb{R}}
\newcommand{\normalN}{\mathcal{N}}
\DeclareMathOperator{\cov}{cov}

\newcommand{\by}{\mathbf{y}}
\newcommand{\bx}{\mathbf{x}}
\newcommand{\bZ}{\mathbf{Z}}
\newcommand{\grm}{\mathbf{K}}
\newcommand{\id}{\mathbf{I}}

\newcommand{\randeff}{u}
\newcommand{\randcoef}{\alpha}
\newcommand{\noise}{\epsilon}

\newcommand{\true}{\beta}
\newcommand{\trueh}{\delta}
\newcommand{\truevar}{\sigma}
\newcommand{\truecov}{\mathbf{G}}
\newcommand{\herit}{h^{2}}
\newcommand{\snr}{\eta}
\DeclareMathOperator{\LMM}{LMM}

\newcommand{\est}{\wh{\true}}
\newcommand{\esth}{\wh{\trueh}}
\newcommand{\estlmm}{\est_{\LMM}}

\newcommand{\sv}{\sigma}
\newcommand{\svmin}{\sv_{\min}}
\newcommand{\svmax}{\sv_{\max}}
\newcommand{\htol}{\gamma}

\newcommand{\mrt}{\textsf{MR-PC(3)}}
\newcommand{\mr}{\textsf{MR}}
\newcommand{\lms}{\textsf{LMM-simple}}
\newcommand{\lmf}{\textsf{LMM-full}}
\newcommand{\lme}{\textsf{LMM-select}}
\newcommand{\lml}{\textsf{LMM-lowRank}}
\newcommand{\lmt}{\textsf{LMM-threshold}}
\newcommand{\lmo}{\textsf{LMM-oracle}}
\newcommand{\pv}{$p$-value}
\newcommand{\lmfc}{\textsf{LMM-full-c}}

\newcommand\blfootnote[1]{%
  \begingroup
  \renewcommand\thefootnote{}\footnote{#1}%
  \addtocounter{footnote}{-1}%
  \endgroup
}

\newcommand{\beginsupplement}{%
        \setcounter{table}{0}
        \renewcommand{\thetable}{S\arabic{table}}%
        \setcounter{figure}{0}
        \renewcommand{\thefigure}{S\arabic{figure}}%
        \setcounter{section}{0}
        \renewcommand{\thesection}{S\arabic{section}}%
     }

\title{Tradeoffs of Linear Mixed Models in Genome-wide Association Studies} 

\author[1]{Haohan Wang}
\author[2]{Bryon Aragam}
\author[1]{Eric P. Xing}
\affil[1]{School of Computer Science, Carnegie Mellon University}
\affil[2]{Booth School of Business, University of Chicago}

\date{}

\begin{document} 

\maketitle 

\begin{abstract}
Motivated by empirical arguments that are well-known from the genome-wide association studies (GWAS) literature, we study the statistical properties of linear mixed models (LMMs) applied to GWAS. First, we study the sensitivity of LMMs to the inclusion of a candidate SNP in the kinship matrix, which is often done in practice to speed up computations.
Our results shed light on the size of the error incurred by including a candidate SNP, providing a justification to this technique in order to trade-off velocity against veracity.
Second, we investigate how mixed models can correct confounders in GWAS, which is widely accepted as an advantage of LMMs over traditional methods. We consider two sources of confounding factors---population stratification and environmental confounding factors---and study how different methods that are commonly used in practice trade-off these two confounding factors differently.
\end{abstract}

\section{Introduction}
\blfootnote{$^\star$ epxing@cs.cmu.edu}
Linear mixed models (LMMs) are a popular tool for identifying genetic associations and are known to be powerful in calibrating the distribution of test statistics raised by confounding factors such as population stratification, family structure, and cryptic relatedness \citep{yu2006unified,kang2008efficient,kang2010variance}. 
A long line of empirical work has demonstrated the advantages of LMMs in genetic studies compared to alternative methods \citep{yu2006unified,kang2008efficient,kang2010variance,zhou2012genome,segura2012efficient,korte2012mixed,svishcheva2012rapid,loh2015efficient}.
As a result, a profusion of methods based on mixed modeling methodology has emerged with the justification that these methods either directly or indirectly correct for ``hidden'' population structure.
In this paper, we show how several methodological tricks used in practice can be justified and provide evidence based on real data to illustrate these results. These results offer practical guidance for researchers using LMMs in genetic studies. 
A central question throughout this paper is: 
\emph{What are the strengths of LMMs in GWAS, and how can this understanding ameliorate the usage of LMMs in different practical situations? }

\begin{table}[tp]
\centering 
\begin{tabular}{lccccc}
\hline
 & \textbf{M} & \textbf{I} & \textbf{P} & \textbf{E} & \textbf{C} \\ \hline
\textsf{Marginal Regression} &  & \checkmark &  &  & $O(pn^2)$ \\
\lms{} \citep{yang2014advantages} & \checkmark &  & \checkmark &  & $O(pn^2 + n^3)$ \\
\lmf{} \citep{yang2014advantages}& \checkmark & \checkmark & \checkmark &  & $O(pn^3)$ \\
\lme{} \citep{listgarten2013fast} & \textbf{?} & \textbf{?} & \textbf{?} &  & $O(pn^2 + sn^2)$ \\
\lml{} \citep{wang2017variable} & \checkmark &  & \checkmark & \checkmark & $O(pn^2 + n^3)$ \\
\lmt{} \citep{tucker2015two} & \checkmark &  & \checkmark & \checkmark & $O(pn^2 + n^3)$ \\ \hline
\end{tabular}
\caption{Comparison of LMM methods. Legends: \textbf{M} indicates whether the method is a multivariate method, \textbf{I} indicates whether the tested SNP is independent on the kinship matrix, \textbf{P} indicates whether the method can correct population stratification, \textbf{E} indicates whether the method can correct environmental confounding factors, \textbf{C} indicates the computational complexity ($n$ denotes the number of samples, $p$ denotes the number of SNPs, and $s$ denotes the number of SNPs selected by \lme{}.). ``\textbf{?}'' denotes that whether \lme{} has that merit depends on how it selects SNPs. }
\label{tab:lmms}
\end{table}

Motivated by this main question, our goal in this paper is to better understand the role of population structure in the success of LMMs, and how this informs practical aspects of LMM methodology such as constructing the kinship matrix. Table~\ref{tab:lmms} summarizes some of our main conclusions by providing a side-by-side comparison of mixed model methods commonly used in practice. These conclusions are based on a combination of theory, simulations, and real data experiments as follows:
\begin{itemize}[leftmargin=*]
    \item In \cref{sec:RR}, we compare the sensitivity of a vanilla mixed model estimator to the inclusion of the candidate SNP in the construction of the kinship matrix, motivated by the work of \citep{yang2010common}. Our results indicate that this sensitivity depends on the ratio of the number of SNPs and the number of samples.  
    \item In \cref{sec:K}, we discuss the effectiveness of LMMs in correcting confounding factors under different assumptions on the true source(s) of the confounding factors. These results offer direct guidance towards the practical construction of the variance components. 
\end{itemize}

\section{Background}
\label{sec:background}
GWAS aims to understand how our genome is related to traits. As is typical, we assume we are given $n$ samples (i.e. patients in a cohort of a study) along with $p$ SNPs and a trait $\by\in\R^{n}$. In principle, we are interested in testing all $p$ SNPs, out of which, we consider there are $k$ associated SNPs to be identified. 
Our interest in this paper is in the problem identifying significant associations, which is to be contrasted with the related problem of estimating heritability  \citep{de2015genomic,dicker2016maximum,jiang2016high,zhou2017unified,steinsaltz2018statistical,bonnet2018heritability,sun2018heritability,wu2018scalable,sankararaman2019fast,schwartzman2019simple,pazokitoroudi2020efficient}. 
We consider quantitative traits, i.e. $\by$ takes on continuous real values. We expect similar results to hold for case-control studies, e.g. by regressing out the effects of other covariates (e.g. age and gender), leaving out what are effective quantitative traits for GWAS analysis.
For a detailed discussion of the effect of population structure in studying quantitative traits, see \citep{haldar2012effect}.

\subsection{The mixed model}

To test the significance of a particular SNP $\bx\in \{0,1,2\}^{n\times 1}$, the following mixed model is commonly used in practice:
\begin{align}
\by 
= \true\bx + \randeff + \noise,
\quad \randeff\sim\normalN(0,\truevar_{\randeff}^{2}\grm),
\quad \noise\sim\normalN(0,\truevar_{\noise}^{2}\id).
\label{eq:lmm:model}
\end{align}

\noindent
Here, $\true$ represents the effect of the SNP $\bx$ on the trait $\by$, and $\randeff$ is a random effect that represents the effect of the remaining $q$ SNPs. The matrix $\grm\in\R^{n\times n}$ is variously known as the \emph{kinship matrix}, \emph{genetic relatedness matrix}, or \emph{realized relationship matrix}.
Since \eqref{eq:lmm:model} implies that $\mathbf{y} \sim \normalN(\true\bx, \truevar_{\randeff}^2\grm + \truevar_{\noise}^2\id)$, the kinship matrix evidently encodes the variance-covariance structure of the trait $\by$. 
Thus, the success of model \eqref{eq:lmm:model} in genetics is largely predicated on constructing an appropriate kinship matrix, 
which will be discussed in detail in the next section. 
For now, we assume $\grm$ is given and continue to describe the construction of test statistics based on model \eqref{eq:lmm:model}.

In the linear mixed model \eqref{eq:lmm:model}, there are two quantities of primary interest in genetic studies: The coefficient $\true$, which measures the strength of the association between the SNP $\bx$ and the trait $\by$, and the heritability $\herit$, which is defined as
\begin{align}
\herit
=\frac{\truevar_{\randeff}^{2}}{\truevar_{\randeff}^{2} + \truevar_{\noise}^{2}}.
\end{align}
We can rewrite the heritability in terms of the signal-to-noise ratio $\snr = \truevar_{\randeff}^{2}/\truevar_{\noise}^{2}$ by noting that $\herit=\eta/(\eta + 1)$. Thus, it suffices to estimate $\snr$ in order to determine $\herit$. In this paper, for reasons that will become clear in the sequel, we define $\trueh=\snr^{-1}=\truevar_{\noise}^{2}/\truevar_{\randeff}^{2}$, and estimate this quantity instead. 

Our aim is to study the estimation of $\true$, i.e. the effect size. A commonly used estimator  \citep[e.g.,][]{lippert2011fast} of $\true$ in GWAS is
\begin{align}
    \estlmm = \dfrac{\bx^T(\esth\id + \grm)^{-1}\by}{\bx^T(\esth\id + \grm)^{-1}\bx},
    \label{eq:lmm:beta}
\end{align}
where $\esth=\esth(\grm)$ is the REML estimate of $\trueh$ \citep{thompson1962problem}. In this expression, two dependencies are of importance: $\esth$ and $\grm$. 
To emphasize this dependence, for any $s>0$ and $\grm\in\R^{n\times n}$, we define the following:
\begin{align}
    \est(\grm;s)
    := \dfrac{\bx^T(s\id + \grm)^{-1}\by}{\bx^T(s\id + \grm)^{-1}\bx},
    \qquad
    \est(\grm)=\est(\grm;\esth),
    \label{eq:lmm:est}
\end{align}
where again $\esth=\esth(\grm)$ is the REML estimator of $\trueh$.
Ultimately, in GWAS we are interested in testing the hypothesis $\true=0$. 

\subsection{The kinship matrix}
\label{sec:sub:kinship}

A typical choice for the kinship matrix is $\grm=\bZ\bZ^{T}$, where $\bZ \in \{0,1,2\}^{n\times q}$ denotes the remaining $q=p-1$ SNPs (i.e. excluding $\bx$). This can be motivated by considering the following special case of \eqref{eq:lmm:model}:
\begin{align}
\label{eq:lmm:model:Z}
\by 
= \true\bx + \bZ\randcoef + \noise,
\quad \randcoef\sim\normalN(0,\truevar_{\randeff}^{2}\id),
\quad \noise\sim\normalN(0,\truevar_{\noise}^{2}\id).
\end{align}

\noindent
This is a random-effects model where the SNPs $\bZ$ serve as covariates \citep{heckerman2018accounting,lippert2011fast,yang2014advantages}.
Another possibility is to include the SNP of interest $\bx$ in $\grm$, i.e. $\grm=\bZ\bZ^{T} + \bx\bx^{T}$.
\citep{yang2014advantages} discussed the potential pitfalls of including the SNP of interest in the kinship matrix, concluding empirically that it results in inflated $p$-values. Since these works, many alternatives have been proposed (as we will discuss later).
One of the goals of this paper is to understand the tradeoffs of these alternatives for $\grm$. 

There is another interpretation of the model \eqref{eq:lmm:model:Z} worth recalling: This model is equivalent to a generalized Ridge estimator in which the fixed effect $\true$ is not penalized \citep{maldonado2009mixed,heckerman2018accounting}. 
Thus, although this model is effective in controlling false positives when compared to traditional univariate methods such as marginal regression, this is easily understood as a multivariate regression method that regresses out the effects introduced by other SNPs as covariates. This helps to explain why LMMs outperform marginal regression, even in settings where population stratification may not exist \citep{wacholder2002counterpoint,freedman2004assessing,wang2004evaluating}. 

\section{Error Analysis of LMM in GWAS} 
\label{sec:analysis}

\subsection{Error Analysis of Estimation Error of Fixed Effect}
\label{sec:RR}
Recall that $\est(\grm)$ (cf. \eqref{eq:lmm:est}) is the estimator of $\true$ using $\grm$ as the kinship matrix and $\esth(\grm)$ as the REML estimator of $\trueh$. A question posed by \citep{yang2014advantages} asks whether $\grm=\bZ\bZ^{T}$ should be preferred over $\grm=\bZ\bZ^{T}+\bx\bx^{T}$, where $\bx$ is the SNP being tested. In this section, we seek to quantify the sensitivity of $\est(\grm)$ to the inclusion of $\bx$.

For any $s\ge0$, let $\svmin(s)$ and $\svmax(s)$ denote the smallest and largest singular values of $(s\id + \bZ\bZ^{T}/n)^{-1}$, respectively. The following quantity will be pivotal in the sequel: 
\begin{align} 
\htol(\bZ) 
= \dfrac{(\svmax(0) + \hat{\delta}(\bZ\bZ^{T}))(\svmin(0) + \hat{\delta}(\bZ\bZ^{T}))}{\svmax(0) - \svmin(0)}
\label{eq:gamma}
\end{align}
\noindent
The following result shows that $\htol(\bZ)$ measures the sensitivity of LMM estimators to the inclusion of $\bx$:

\begin{theorem}
\label{theorem:rr:main}
Let  $\epsilon>0$ and define $\svmin:=\svmin(\esth(\bZ\bZ^{T}))$ and $\svmax:=\svmax(\esth(\bZ\bZ^{T}))$, and $\Delta := \esth(\bZ\bZ^{T}+\bx\bx^{T}) - \esth(\bZ\bZ^{T})$.  
If $\svmax - (1+\epsilon)\svmin> 0$ and $\Delta<\htol(\bZ)\epsilon$, then 
\begin{align}
\label{eq:rr:main2}
\dfrac{1}{1+\epsilon}
< \dfrac{\est(\bZ\bZ^{T} + \bx\bx^{T})}{\est(\bZ\bZ^{T})} 
< 1+\epsilon.
\end{align}
\end{theorem}
\noindent
A similar result holds for the standardized coefficients; see Appendix~S.1 for details.

We can interpret this result as follows: As long as the corresponding estimates of $\trueh$ are sufficiently close, it does not make much of a difference whether or not the SNP $\bx$ is included in the kinship matrix. Exactly how much misspecification is allowed is quantified by the requirement that $\Delta<\htol(\bZ)\epsilon$. Evidently, the scaling factor $\htol(\bZ)$
is crucial here (see Theorem~\ref{thm:htol:asymp} below for more on this quantity). The larger $\htol(\bZ)$ is, the more misspecification is allowed. 
This has the following interesting implications:

\begin{itemize}[leftmargin=*]
    \item When studying a problem with high heritability, the difference between $\esth(\bZ\bZ^{T})$ and $\esth(\bZ\bZ^{T}+\bx\bx^{T})$ may not be negligible, while for low heritability problems, this difference can generally be ignored.
    This is because the bigger the heritability is, the smaller $\delta$ is, the smaller the estimated $\delta$ is (according to \citep{jiang2016high}, the estimated heritability is biased up to a scaling factor), thus the smaller $\Delta$ is allowed. 
    \item Also, the quantity $\svmax(0)^{-1} - \svmin(0)^{-1}$ is related to the population structure of the covariance matrix. For example, 
    when the difference between intra-population covariance and inter-population covariance is more pronounced, the difference between the largest singular value and the smallest singular value is smaller. 
\end{itemize}

We can provide a more concrete description of $\htol(\bZ)$ as $n,p\to\infty$, assuming
$\bZ$ and $\bx$ consist of i.i.d draws from some fixed (not necessarily Gaussian) distribution. This allows us to provide deeper insight into the properties of the estimator $\est(\grm)$ in a simple setting. Previous work also suggested that the properties of the covariance matrix constructed by matrices sampled from Gaussian distribution can be helpful in studying the properties of variance component constructed from the SNP matrix \citep{patterson2006population}.

\begin{theorem}
\label{thm:htol:asymp}
Assume $p/n\to\zeta$, $k/p\to\omega$, where $k$ denotes the number of associated SNPs, 
and let $[\bZ, \bx]$ be a matrix of i.i.d draws with zero mean and standard deviation of 1. Then
\begin{align*}
    \htol(\bZ) 
    &\overset{p}{\longrightarrow} \frac{(\zeta + \frac{\trueh}{\omega})^2 + \frac{2\trueh}{\omega} - 2\zeta + 1}{4\sqrt{\zeta}}
    = \frac{\trueh ^2+2\omega \trueh  (\zeta +1)  +\omega ^2(\zeta -1)^2 }{4\omega^2\sqrt{\zeta}}.
\end{align*}
\end{theorem}

\noindent
According to Theorem~\ref{theorem:rr:main}, the larger $\htol(\bZ)$ is, the more tolerance there is in specifying $\grm$. Theorem~\ref{thm:htol:asymp} indicates that---asymptotically---$\htol(\bZ)$ blows up if either $\zeta \to 0$ or $\zeta \to \infty$. This has the following interpretations:
\begin{itemize}[leftmargin=*]
    \item When $\zeta \to 0$, the problem of identifying associated SNPs becomes a simple problem as the sample size is large relative to the number of typed SNPs. In this case, even marginal regression may be able to identify the associated SNPs, the misspecification of $\trueh$ should not have a significant effect. 
    \item When $\zeta \to \infty$, the problem of identifying associated SNPs becomes a hard problem as we have many more SNPs to inspect than the number of samples. In this case, the effect of including the single SNP $\bx$ in $\grm$ will be washed out by the remaining $q\gg n$ SNPs.
\end{itemize}

\subsection{Error Analysis of Approximation Error of Variance Component}
\label{sec:K}
This section studies if and how mixed models can correct different sources of confounding in GWAS. We first formalize this problem by decomposing the covariance of the phenotypes into two components as follows: 
\begin{align*}
    \cov(\by) = \truevar_{\randeff}^{2}\truecov + \truevar_{\noise}^{2}\id.
\end{align*}
The first component $\truevar_{\randeff}^{2}\truecov$ is the variance introduced from the confounding factors
and the second component $\truevar_{\noise}^{2}\id$ is the residual variance. 
We consider two possible sources of confounding factors: 
\begin{itemize}[leftmargin=*]
    \item Population stratification \citep{devlin1999genomic,devlin2001genomic,bacanu2002association}: 
    That is, the confounding due to the unobserved population structure in the data. 
    In our work, the population structure is represented by the untyped, associated SNPs that share similar allele frequencies with typed, non-associated SNPs.
    We model this source of confounding by letting $\truecov=\truecov_{\textnormal{pop}} := \mathbf{E}\mathbf{E}^T/l$ where $\mathbf{E} \in \{0,1,2\}^{n\times l} $ is another, unobserved SNP array from the same individuals. 
    In particular, this implies that $\mathbf{E}$ has the same distribution $\mathbf{Z}$.

    \item Environmental confounding factors \citep{vilhjalmsson2013nature}: That is, confounding due to the existence of heterogeneous subpopulations, e.g. as defined in \citep{siva20081000}.
    We model this source of confounding by letting $\truecov=\truecov_{\textnormal{env}}(\rho)$, where $\truecov_{\textnormal{env}}(\rho)$ is a block-diagonal square matrix given by
    \begin{align*}
    [\truecov_{\textnormal{env}}(\rho)]_{ij}
    = \begin{cases}
    \rho & \text{$(i,j)$ in the same subpopulation,} \\
    0 & \text{otherwise.}
    \end{cases}
    \end{align*}
    In other words, if we use $\mathcal{C}$ to denote the set whose elements are pairs of subjects that are from the same population ($(i,j)\in \mathcal{C}$ means that $i$, $j$ are in the same subpopulation), we have $[\truecov_{\textnormal{env}}(\rho)]_{ij}=\rho$ as long as $(i,j)\in\mathcal{C}$.
\end{itemize}
Later, our results suggest that LMM naturally approximates $\truecov_{\textnormal{pop}}$ well, but a more deliberately designed kinship matrix can help LMM approximate $\truecov_{\textnormal{env}}$ more accurately. 
Notice that in the discussion of confounding factors, we will define $\mathbf{K}=\bZ\bZ^T/q$ for convenience, which only differs from the previous case in a constant. 

\subsubsection{Approximating Population Stratification Variance Component}
\label{sec:K:pop}

We follow the basic set-up of the population stratification in quantitative traits as argued by \citep{bacanu2002association}. 
Since $\mathbf{E}$ has the same distribution $\mathbf{Z}$, $\truecov_{\textnormal{pop}} = \mathbf{E}\mathbf{E}^T/l$ is also distributed similarly to $\mathbf{K} = \mathbf{Z}\mathbf{Z}^T/q$, and we have:
\begin{prop}
Assuming $\mathbb{E}[\truecov_{\textnormal{pop}}] = \mathbb{E}[\mathbf{K}]$, we have
\begin{align*}
    \mathbb{P}(\Vert\mathbf{K} - \mathbb{E}[\truecov_{\textnormal{pop}}]\Vert_2 \geq t)
\leq 2n\exp(-\dfrac{3t^2q}{96n^2 + 8nt}).
\end{align*}
\label{lemma:k:pops}
\end{prop}
The proof of this bound is a straightforward application of matrix concentration inequalities.
The result matches the intuition: Increasing $q$ (i.e. the number of SNPs) improves estimation, whereas increasing $n$ (i.e. the number of individuals) makes estimation more difficult. This is because increasing $n$ increases the effective number of parameters to be estimated (i.e. since $\truecov_{\textnormal{pop}}$ is an $n\times n$ matrix), whereas increasing $q$ does not affect the number of parameters.

\subsubsection{Approximating Environmental Confounding Factor Variance Component}
\label{sec:K:en}

We compare three different methods of building the variance component in LMM and how well they approximate the simplified covariance structure $\truecov_{\textnormal{env}}(\rho)$. 
We will focus on the case when the candidate SNP is excluded from the kinship matrix; the case when the candidate SNP is included is similar and in fact, leads to the same conclusions.
\begin{itemize}[leftmargin=*]
    \item $\mathbf{K} = \mathbf{Z}\mathbf{Z}^T/q$: The vanilla LMM, which is probably the most popular way of constructing the variance component. 
    \item $\tilde{\mathbf{K}}$: Low-rank approximation of $\mathbf{K}$ through truncated singular value decomposition (TSVD) approximation. 
    The TSVD uses a low-rank approximation to $\mathbf{K}$ by selecting significant singular values according to some criteria \citep{hoffman2013correcting,wang2017variable}.
    \item $\grave{\mathbf{K}}(\tau)$: Thresholded kinship matrix defined as $\grave{\mathbf{K}}(\tau)_{i,j} = \mathbf{K}_{i,j}$ if $\mathbf{K}_{i,j} > \tau$, and $\grave{\mathbf{K}}(\tau)_{i,j} = 0 $, otherwise, where $i,j$ are indices of $\mathbf{K}$ \citep{tucker2015two}.
\end{itemize}

Existing work suggests that the vanilla LMM is outperformed by leveraging either the low-rank or thresholded kinship matrices defined above. To check this, we can ask which constructions lead to better approximations of the true covariance, which we are assuming to be $\truecov_{\textnormal{env}}(\rho)$ for simplicity. Obviously more complicated structures arise in real data, however, this simple assumption allows for some immediate insights into the general problem. For example, the following result establishes a necessary and sufficient condition for the approximation error of $\mathbf{K}$ to dominate the approximation error of both $\tilde{\mathbf{K}}$ and $\grave{\mathbf{K}}(\tau)$:

\begin{prop}
\label{theorem:K:main}
Consider approximating
$\truecov_{\textnormal{env}}(\rho)$ with one of the three kinship matrices defined above. Then, for an optimally tuned threshold $\tau^\star$, we have 
\begin{align*}
    \Vert\mathbb{E}[\grave{\mathbf{K}}(\tau^\star)] - \mathbf{G}_{\textnormal{env}}(\rho)\Vert_F^2 \leq 
    \Vert\mathbb{E}[\tilde{\mathbf{K}}] - \mathbf{G}_{\textnormal{env}}(\rho)\Vert_F^2 \leq 
    \Vert\mathbb{E}[\mathbf{K}] - \mathbf{G}_{\textnormal{env}}(\rho)\Vert_F^2,
\end{align*}
if and only if 
\begin{align}
\begin{split}
    (\mathbb{E}[\tilde{\mathbf{K}}_{i,j|(i,j) \not\in \mathcal{C}}])^2 - (\mathbb{E}[\mathbf{K}_{i,j|(i,j) \not\in \mathcal{C}}])^2
    &\leq \\
    (\mathbb{E}[\mathbf{K}_{i,j|(i,j) \in \mathcal{C}}]-\rho)^2 - 
    (\mathbb{E}[\tilde{\mathbf{K}}_{i,j|(i,j) \in \mathcal{C}}]-\rho)^2
    &\leq 
    (\mathbb{E}[\tilde{\mathbf{K}}_{i,j|(i,j) \not\in \mathcal{C}}])^2.
\end{split}
\label{eq:therorem:K:assumption}
\end{align}
\end{prop}

Thus, to check when the performance of $\grave{\mathbf{K}}(\tau^\star)$ is better than $\tilde{\mathbf{K}}$ (and similarly $\tilde{\mathbf{K}}$ is better than $\mathbf{K}$), it suffices to check the condition \eqref{eq:therorem:K:assumption}, 
and we validated that this condition is likely to hold in practice with simulations in the appendix. 

Proposition~\ref{theorem:K:main} 
guides us about the choice of the construction of kinship matrix when the sources of confounding factors in consideration are environmental confounding factors. 
As the analysis suggests, the two-variance-component method \citep{tucker2015two} outperforms the truncated-rank methods \citep{hoffman2013correcting,wang2017variable}, which outperform the vanilla set-up.
Of course, the determination of $\tau^\star$ for the two-variance-component method can be tricky in practice, which highlights some of the practical challenges in using this method.
The assumption we rely on can be intuitively understood as the tendency of truncated rank SVD reconstruction method of approximating the original matrix by emphasizing the ``major'' components (i.e., the information within the block diagonal) and ignoring the ``minor'' components (i.e., the information off the block diagonal). 

\section{Simulations}
\label{sec:results}
\input{secs/simulations}

\section{Analysis on Genetic Data of Alzheimer's Disease, Drug Abuse, and Alcoholism}
\label{sec:real}

Further, 
we compare the six methods  (i.e., \mrt{}, 
\lms{}, \lmf{}, \lme{}, \lml{}, \lmt{}) on three real-world GWAS data sets. 
To account for the heterogeneous nature of the genetic data, we use \mrt{} to replace \mr{} by regressing three principle components out of the phenotype as a baseline method to control population structure.
Further, as these real data sets have over 200k SNPs, it is computationally challenging to run \lmf{} on these data sets. 
Fortunately, with the techniques introduced by \citep{listgarten2012improved}, we only need to construct a new kinship matrix every time the testing proceeds to a new chromosome. We refer to this method as \lmfc{}. 
Also, we choose $\tau=0.05$ as suggested by \citep{tucker2015two} for real data in \lmt{}.

\begin{figure}
\centering
\quad
\includegraphics[width=0.9\textwidth]{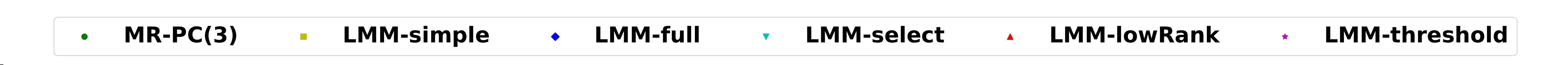}
\quad \\
\subfloat[DA (top 500 SNPs)]{\includegraphics[width=0.3\textwidth]{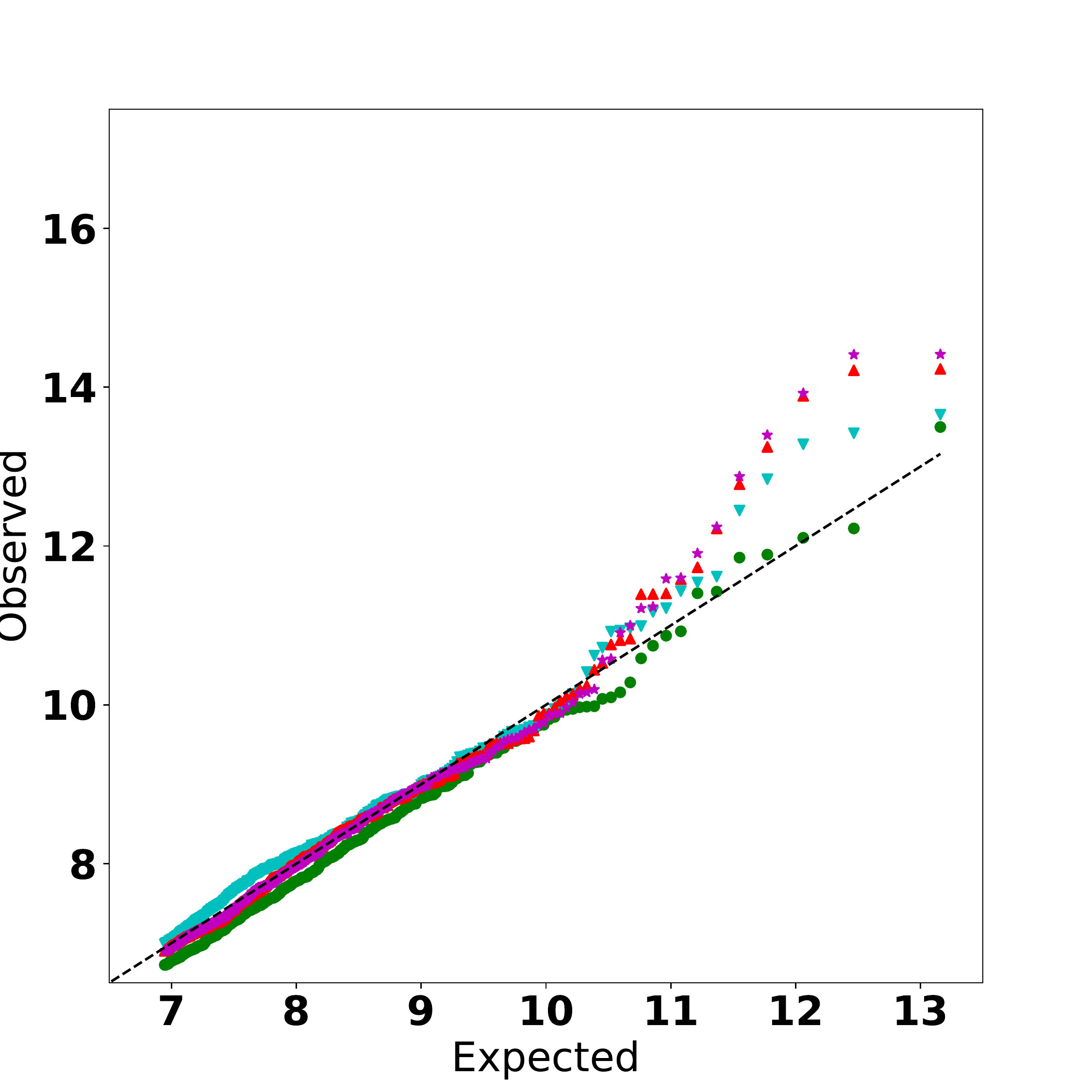}}
\quad
\subfloat[ALC (top 500 SNPs)]{\includegraphics[width=0.3\textwidth]{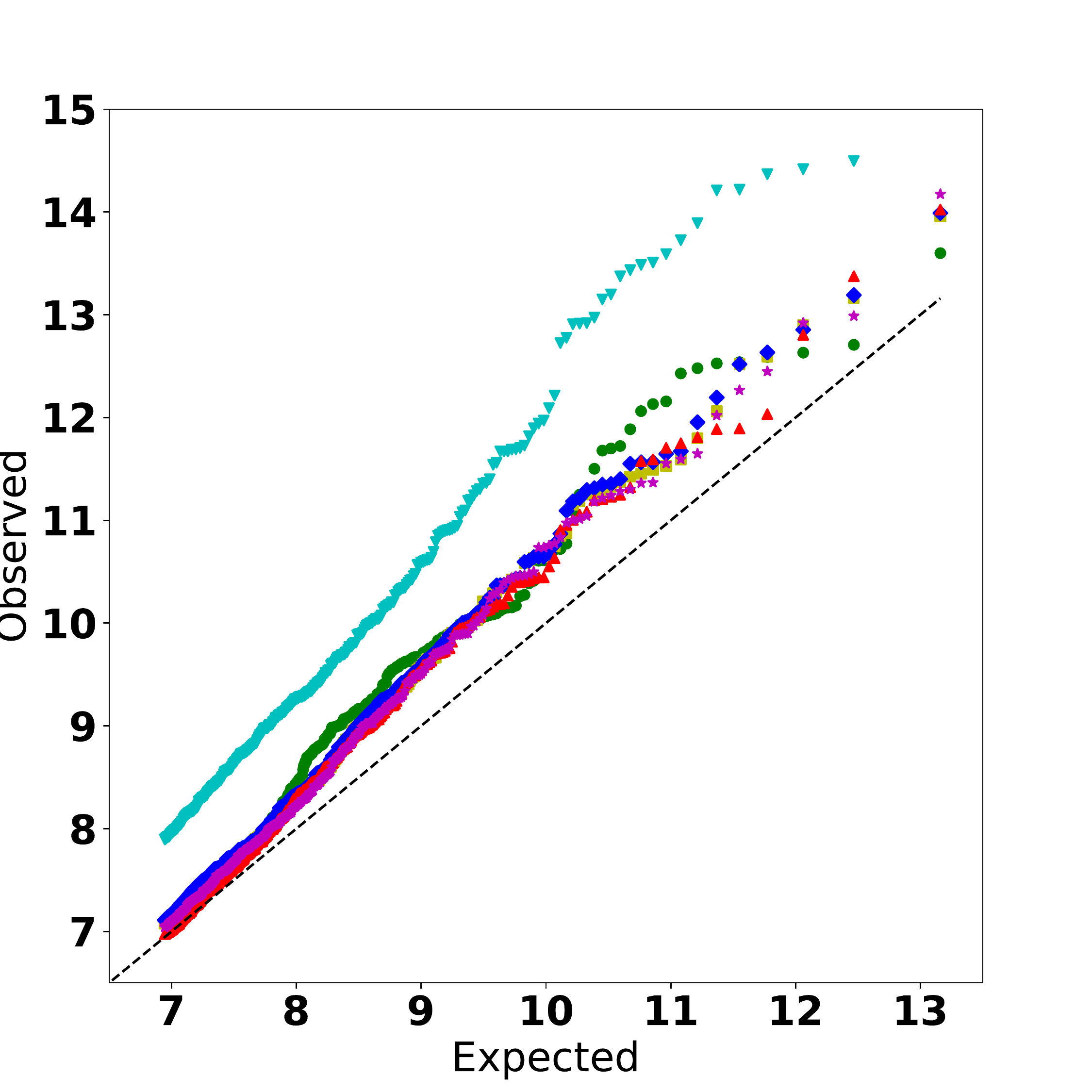}}
\quad
\subfloat[AD (top 500 SNPs)]{\includegraphics[width=0.3\textwidth]{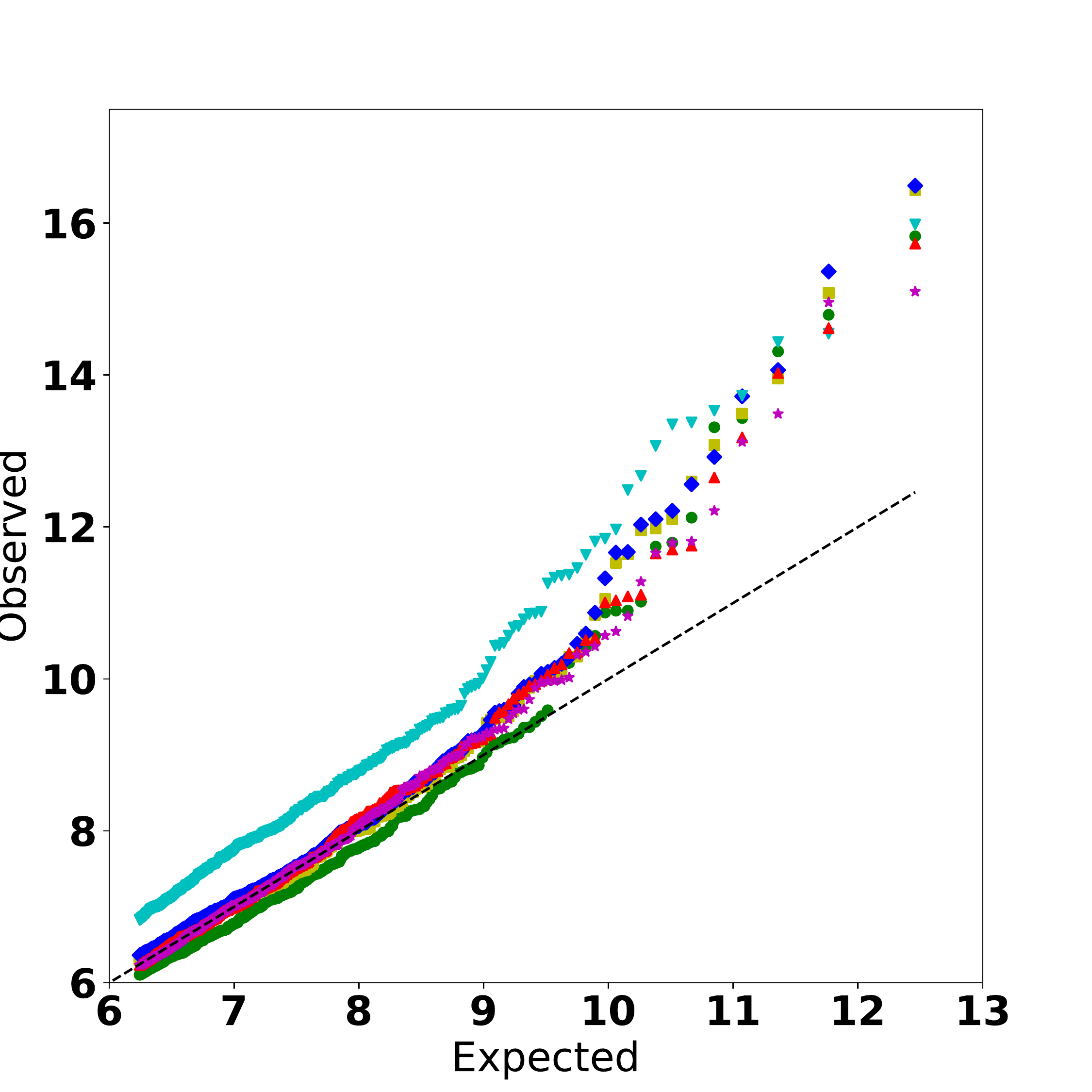}}\\
\subfloat[DA (evenly sampled)]{\includegraphics[width=0.3\textwidth]{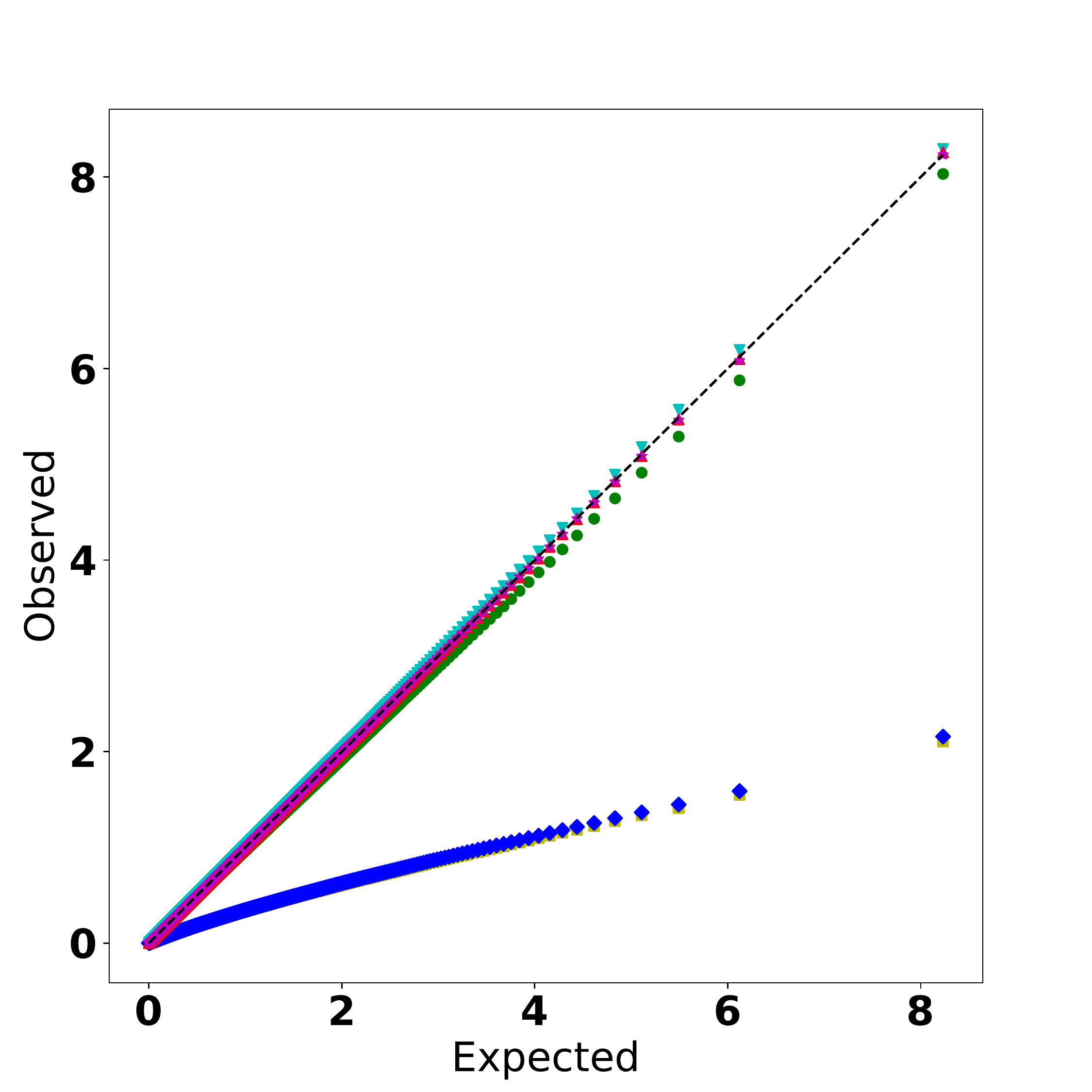}}
\quad
\subfloat[ALC (evenly sampled)]{\includegraphics[width=0.3\textwidth]{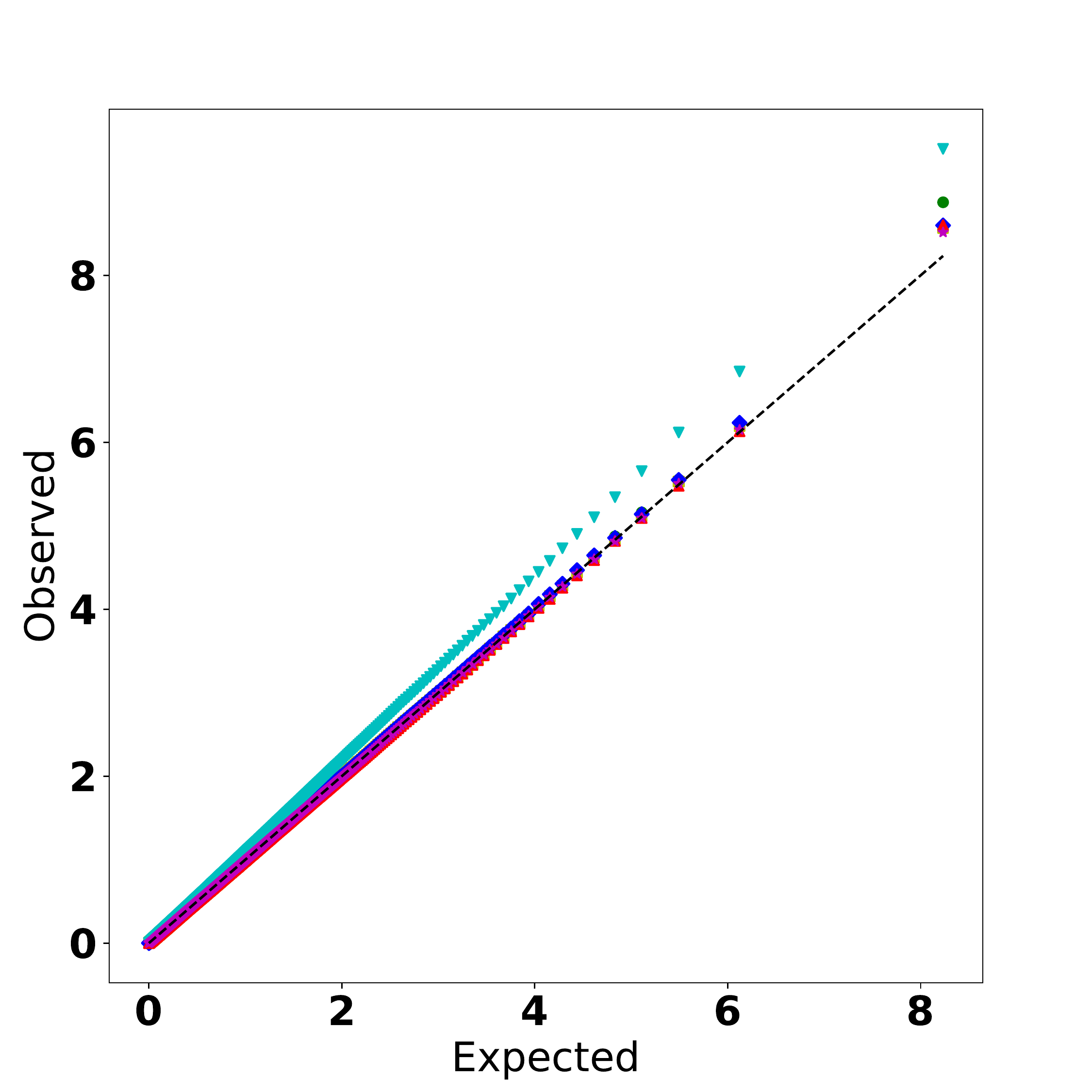}}
\quad
\subfloat[AD (evenly sampled)]{\includegraphics[width=0.3\textwidth]{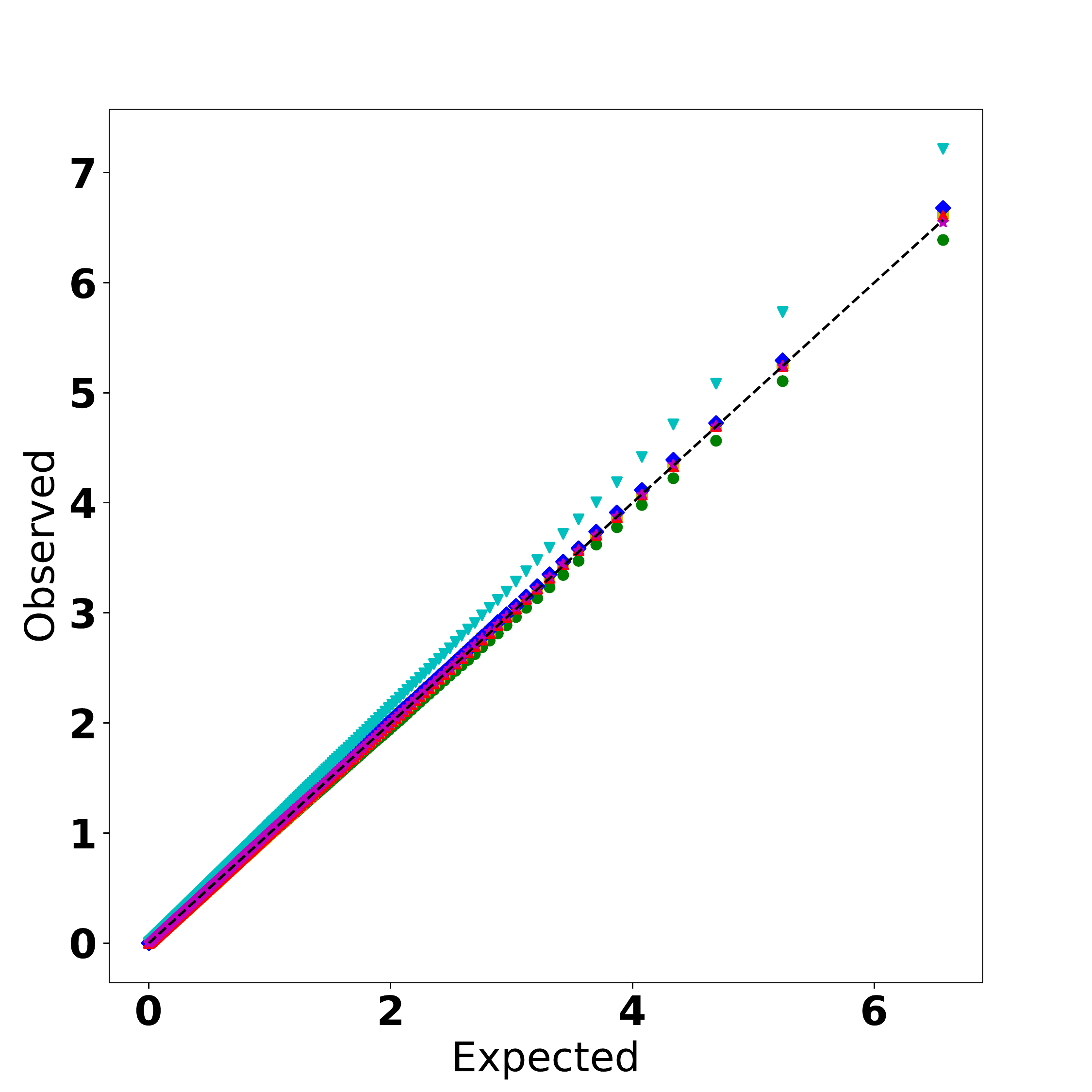}}\\
\caption{QQ-plot of the comparison methods over three data sets (DA: drug abuse; ALC: Alcoholism; AD: Alzheimer's Disease). The top row figures illustrate the most significant 500 SNPs of each data set (some dots are not shown because they significantly deviate from the expectation). The bottom row figures show the over all distribution of \pv s of all the SNPs, evenly sampled one every 1000 SNP.}
\label{fig:real}
\end{figure}

\subsection{Data Sets:}
We consider three datasets:
1) The \textbf{Drug Abuse (DA)} dataset was collected by the CEDAR Center at the University of Pittsburgh\footnote{http://www.pitt.edu/~cedar/}, with 359 patients (153 cases and 206 controls) and roughly 519k SNPs. 
2) The \textbf{Alcoholism (ALC)} dataset was collected by the CEDAR Center at the University of Pittsburgh with 383 samples (78 cases and 305 controls) and also roughly 519k SNPs. 
3) The \textbf{Alzheimer's Disease (AD)} dataset was collected from the Alzheimer’s Disease Neuroimaging Initiative (ADNI)\footnote{http://adni.loni.usc.edu/}. We only consider the individuals diagnosed with AD or normal controls. There are 477 individuals (188 cases and 289 controls). There are roughly 257k SNPs. 
To the best of our knowledge, the subjects of DA and ALC data are collected with family structure in consideration, while the subjects of AD data are not related. 

Although there is some overlap between the control group of DA and ALC datasets, there is no overlap between the case groups.
We also exclude the SNPs on the X-chromosome following suggestions of previous studies \citep{bertram2008genome}.
Missing values have been imputed as the mode of the corresponding SNPs.

\subsection{Results:}
We tested the six methods on these three datasets with the resulting QQ-plots shown in Figure~\ref{fig:real}. 
We plot for both the top 500 most significant SNPs (top row) and the SNPs evenly sampled one every 1000 SNP, sorted by $p$-value (bottom row). 
These results interestingly reveal different properties of these methods. 

\paragraph{All SNPs (bottom row):} 
Interestingly, \pv s from \lme{} inflate with a large margin, suggesting that conditioning on a subset of the most associative SNPs is insufficient to control the inflation of all the SNPs.
The \pv s from \lml{} and \lmt{} stay closely to the expectation.
On the other hand, \pv s from \lms{} and \lmfc{} are significantly below the expectation for the DA data set. 
This difference is impressive because DA is widely considered to be related to environmental factors such as access to drugs and neighborhood influence \citep{jedrzejczak2005family,mennis2016risky}. 
This suggests that the differences between \lms{}, \lmfc{} and \lml{}, \lmt{} are primarily due to the effects in controlling the environmental confounding factors, similar to what our simulation experiments suggest. 
Also, we notice the \pv s reported by \lms{} and \lmfc{} almost always overlap with each other, suggesting whether conditioning on the candidate has a negligible overall impact, 
which again aligns with our discussions in Theorems~\ref{theorem:rr:main} and~\ref{thm:htol:asymp} since for these datasets $p\gg n$.

\paragraph{Top SNPs (top row):}
For DA, results from \lms{} and \lmfc{} cannot be plotted without a significant twist of the figure layout, so we discard these points. The results of \lms{} and \lmfc{} for DA are not comparable to the rest methods. 
Generally speaking, we can still see the strong inflation of \lme{}. However, for DA, \pv s from \lme{} are the closest to the expectation. We believe this is because these top SNPs were picked up by the first phase of \lme{}, thus \lme{} conditioned on these SNPs during testing. This property also helps explain that the results of \lme{} tend to fluctuate the most for the top SNPs: the fluctuation may depend on whether \lme{} has selected certain SNPs to condition on (therefore depend on how \lme{} select the SNPs). 
Again, we can see that \lms{} and \lmfc{} behave almost exactly the same. 
Also, \lml{} and \lmt{} behave quite similarly. 

\subsection{Discussion}
In summary of our experiments, we have:
\begin{itemize}[leftmargin=*]
    \item \lms{} and \lmfc{} behave surprisingly close even in the real-world data ($p\gg n$), supporting the discussion of Theorem~\ref{theorem:rr:main} and ~\ref{thm:htol:asymp}. 
    \item \lml{} and \lmt{} perform quite similarly, implying that $\tau=0.05$, as suggested by \citep{tucker2015two}, is a reasonably good but maybe not optimal threshold for real-world data, according to the discussion of Proposition~\ref{theorem:K:main}. 
    \item For AD and ALC, \lms{}, \lmfc{}, \lml{}, and \lmt{} all behave similarly, but for DA, \lml{} and \lmt{} are significantly better than \lms{} and \lmfc{}, suggesting the importance of handling environmental confounding factors when such factors exist. On the other hand, when such factors do not exist, it is unnecessary to employ methods that can correct environmental confounding factors. 
\end{itemize}

We have summarized our discussions in Table~\ref{tab:lmms}, where ``\checkmark''d denotes the method has the marked merit, and ``\textbf{?}'' denotes that whether \lme{} has that merit depends on how \lme{} selects the SNPs to construct the variance component. 

\input{secs/discussion}

\section{Conclusion}
\label{sec:con}
In this paper, we focused on the tradeoffs of LMMs when applied to GWAS. We first studied whether it is crucial to include the candidate SNP into the construction of the kinship matrix, and then we study the effectiveness of estimating the variance component of confounding factors. 

Finally, we have open-sourced a Python package\footnote{\href{https://github.com/HaohanWang/LMM-Python}{https://github.com/HaohanWang/LMM-Python}} that includes all the major LMMs that are discussed throughout this paper. We hope that this tool will prove useful both for practitioners using these methods on real data, and also to enable future methodological research on mixed models and GWAS.

\subsection*{Acknowledgments}
The authors would like to thank Steven Knopf and Dr. Michael M. Vanyukov from University of Pittsburgh for instructions in using the Alcoholism and Drug Abuse data.
This work is supported by the National Institutes of Health grants R01-GM093156 and P30-DA035778. 
Data collection and sharing for this project was funded by the Alzheimer's Disease Neuroimaging Initiative (ADNI) (National Institutes of Health Grant U01 AG024904) and DOD ADNI (Department of Defense award number W81XWH-12-2-0012).

\bibliographystyle{abbrvnat}
\bibliography{ref}

\newpage 
\beginsupplement
\input{secs/appendix}


\end{document}

%% file: secs/simulations.tex

We conduct simulation experiments to verify these results. We compare the following methods: 
\begin{itemize}
    \item \mr{} (Marginal Regression) serves as baseline reference for all other comparisons. 
    \item \lms{}: One of the methods discussed in Section~\ref{sec:RR}, i.e. the one with $\est(\bZ\bZ^{T}+\bx\bx^{T})$
    \item \lmf{}: One of the methods discussed in Section~\ref{sec:RR}, i.e. the one with $\est(\bZ\bZ^{T})$
    \item \lme{} \citep{listgarten2013fast} constructs the variance component with some identified SNPs that are highly associated with phenotypes.
    \item \lml{}: Method $\bar{\mathbf{K}}$ discussed in Section~\ref{sec:K}. The rank was automatically selected with method in \citep{wang2017variable}. 
    \item \lmt{}: Method $\grave{\mathbf{K}}(\tau)$ discussed in Section~\ref{sec:K}. In the simulation experiments, we notice that $\tau=0.001$ is a reasonable choice. 
    \item \lmo{}: We use the oracle population partition as $\mathbf{K}$ in LMM, corresponding to the situation where we have the oracle $\tau$ in $\grave{\mathbf{K}}(\tau)$.
\end{itemize}



The choice of these methods used in the sequel reflects the two different questions studied in the previous section.
Specifically, we use \mr{}, \lms{}, \lmf{}, and \lme{} to verify the discussion of  Theorem~\ref{theorem:rr:main} and~\ref{thm:htol:asymp} in \cref{sec:exp:mr} and \cref{sec:exp:exc}; we use \mr{}, \lms{}, \lml{}, and \lmt{} to verify the discussion of Proposition~\ref{theorem:K:main} in \cref{sec:exp:pop} and \cref{sec:exp:env}. 
For each experiment, we consider the SNPs identified by the methods with the $p$-values less than 0.05 after the $p$-value is corrected by Benjamini-Hochberg (BH) procedure \citep{benjamini1995controlling}. We report both precision and recall of these identified SNPs. 


\begin{figure}
    \centering
    \includegraphics[width=1.0\textwidth]{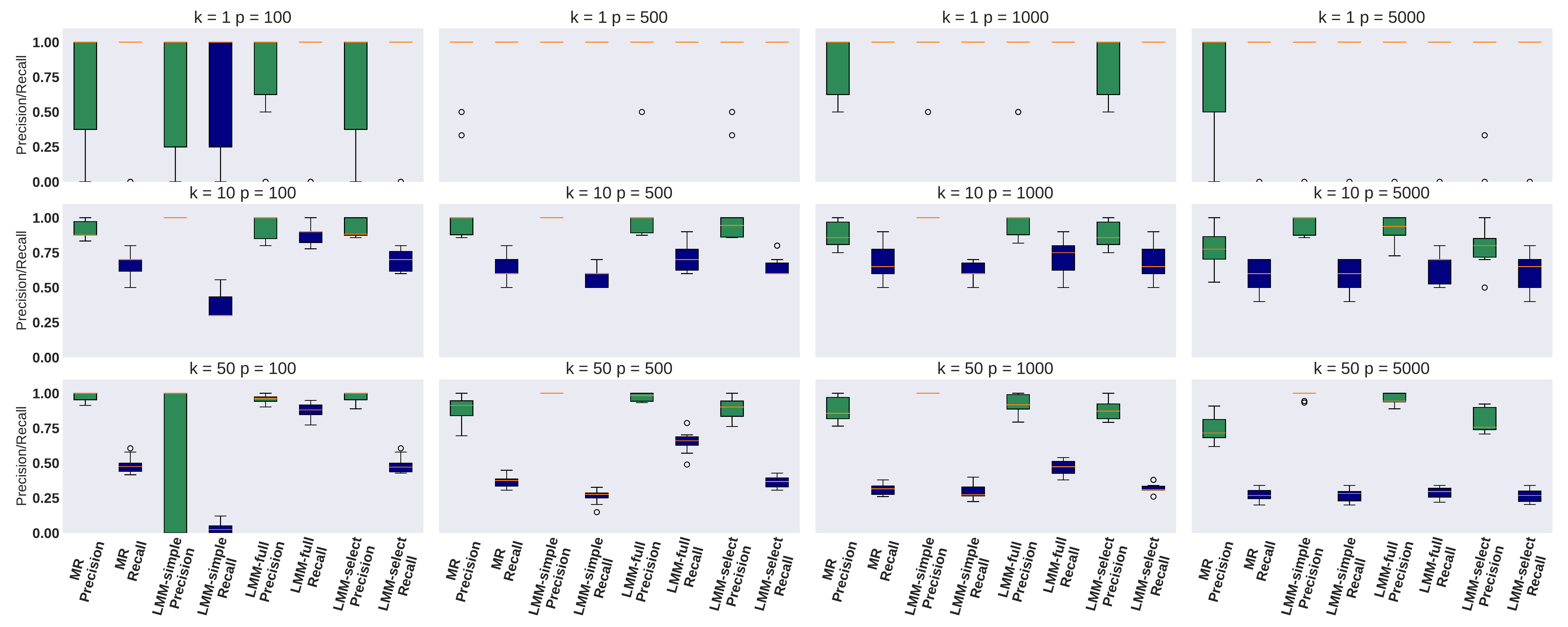}
    \caption{LMM is superior to MR in most cases when there are multiple associated SNPs even when there are no confounding factors in the data.}
    \label{fig:exp1}
\end{figure}

\subsection{SNP data without Population Structure}
\label{sec:exp:mr}

\quad 

\textbf{Data Generation:} We use SimuPop \citep{peng2005simupop} to generate SNP data. 
Due to the high $O(np^3)$ computational complexity of \lmf{}, we only consider relatively small datasets with $p\in\{100, 500, 1000, 5000\}$ SNPs and the minor allele frequency set to $0.1$. 
$n$ is set to be 500. 
There are $k\in\{1, 10, 100\}$ SNPs associated with the phenotype, with the effect sizes (for associated SNP $i$) sampled as : 
$\mathbf{\beta}_i = W + Z$, $W\in\{-1,1\}$, $W\sim \text{Ber}(1/2)$, $Z\sim N(0, 1)$ ($\mathbf{\beta}$ in all our simulations are sampled in this same manner).
Finally, the phenotype is sampled as:
$\mathbf{y} = \mathbf{X}\mathbf{\beta} + \mathbf{\epsilon}$, $\mathbf{\epsilon} \sim N(0, \mathbf{I})$. 


\textbf{Association Study:} 
Figure~\ref{fig:exp1} shows both the mean precision (in green) and mean recall (in blue) of identifying the association SNP over 10 random seeds across the experiment settings.

\emph{\lms{} vs. \mr{}:} 
With only one associated SNP ($k=1$), \mr{} and \lms{} behave similarly. 
However, as the number of associated SNPs (i.e., $k$) increases, \lms{} starts to outperform \mr{}, especially in high-dimensions. 
This confirms what is already known: \lms{} is essentially a multiple regression method \citep{maldonado2009mixed,heckerman2018accounting}. 
Also, we notice that \lms{} has higher precision compared to \mr{},
however, in many cases, \lms{} does not show a comparable recall with \mr{}, which is likely because the candidate SNP is included in the construction of the variance component.

\emph{\lms{} vs. \lmf{}:} Across these four methods, \lms{} behaves most closely to \lmf{}. However, differences can still be observed as a trade-off: \lmf{} in general yields inferior precision but superior recall in comparison to \lms{}. 
This behavior links to the fact that \lmf{} does not condition on the candidate SNP, thus may lead to smaller $p$-values for the candidate SNP. 
Thus \lmf{} tends to report more significant SNPs, naturally leading to lower precision but higher recall. 
Also, as $p$ increases, we notice that 
the differences between \lms{} and \lmf{} diminish. 
\lms{} is even marginally better than \lmf{} when $p=5000$. 
This aligns well with our understanding in Theorems~\ref{theorem:rr:main}
and~\ref{thm:htol:asymp}: 
when $p \gg n$ (in our case $p/n=10$), 
the specific choice of $\grm$ has less effect, thus
the difference between 
\lms{} and \lmf{} is negligible.


\emph{\lms{}, \lmf{}, vs. \lme{}}: In our implementation, \lme{} conditions on the set of SNPs identified as trait-associted SNPs by marginal regression. 
Therefore, the performance of \lme{} crucially depends on this set. 
Specifically, as the model tests the candidate SNP one by one, the performance falls into two cases: 1) the candidate SNP is within the identified subset (so \lme{} is similar to \lms{}) and 2) the candidate SNP is not within the identified subset (so \lme{} is similar to \lmf{}).
As a result, as associated SNPs are more likely to be identified by the linear regression, \lme{} behaves more closely to \lms{} for associated SNPs, and more similar to \lmf{} for non-associated SNPs. 
Our simulation suggests that \lme{} has a trade-off with \lms{} in terms of precision and recall, but it is occasionally strictly worse than \lmf{}.


\begin{figure}
    \centering
    \includegraphics[width=1.0\textwidth]{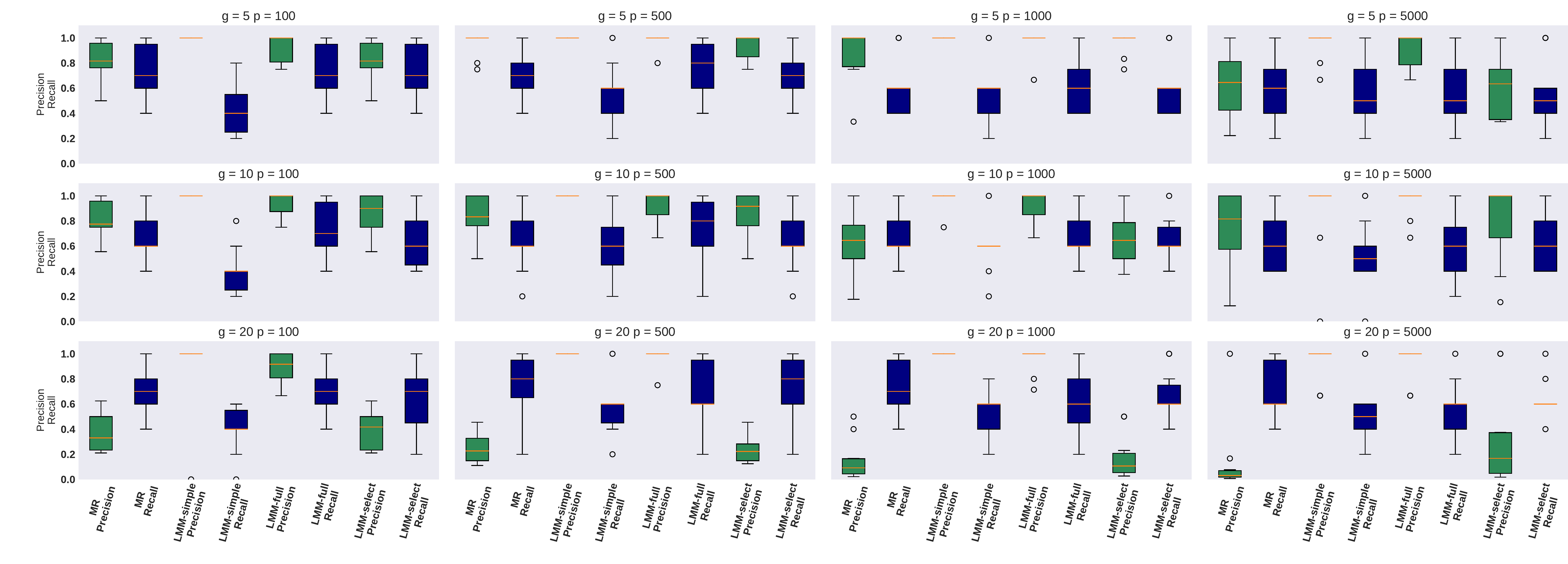}
    \caption{The performance of precision and recall for different LMMs when the data has population structure, but no confounding factors are introduced. }
    \label{fig:exp2}
\end{figure}

\subsection{SNP data with Population Structure}
\label{sec:exp:exc}
Although our analysis regarding the sensitivity of $\hat{\beta}$ does not depend on the underlying population structure, for completeness we ran experiments with population structure. 
Overall, we observe similar conclusions. 

\textbf{Data Generation:} 
During the generation, $n=500$ samples are randomly split into 5 different populations, samples within each population reproduce for $g$ generations and start to split into 5 sub-populations at the $8$\textsuperscript{th} generation (if $g>8$). 
Also, each sample has a small probability (i.e., 0.01) of migrating into other populations during the reproduction process. 
We fixed $k=10$ and experimented with the settings $g\in\{5, 10, 20\}$ with $p \in \{100, 500, 1000, 5000\}$. We sampled the data with 10 different random seeds. 


\textbf{Association Study:} 
Again, Figure~\ref{fig:exp2} shows both the mean precision and mean recall of identifying the association SNP over 10 random seeds across the experiment settings. 

\emph{\lms{} vs. \mr{}:} 
Compared to the experiments without population structure, \mr{} shows a noticeable drop in performance (middle row of Figure~\ref{fig:exp1} and Figure~\ref{fig:exp2} when $k=10$ in both cases), while the performance \lms{} is barely changed. 
This is consistent with the well-known observation that mixed models outperform \mr{} in the presence of population structure
\citep[e.g.,][]{yu2006unified,kang2008efficient,kang2010variance}.

\emph{\lms{} vs. \lmf{}:} The differences between \lms{} and \lmf{} are similar to the previous case: \lmf{} reports inferior precision but superior recall in general. 
Further, the comparison of the middle row of Figure~\ref{fig:exp1} and Figure~\ref{fig:exp2} suggests that the changes of performances of \lms{} and \lmf{} are marginal when the population structure is introduced.  
Similar to the observation in Section~\ref{sec:exp:mr}
that support Theorem~\ref{theorem:rr:main} and~\ref{thm:htol:asymp}, 
when $p$ is large ($p=5000$, thus $p/n=10$), 
the difference between \lms{} and \lmf{} seems negligible. 
Interestingly, 
we notice that \lms{} even outperforms \lmf{} in general when $p=5000$. 

\begin{figure}
    \centering
    \includegraphics[width=1.0\textwidth]{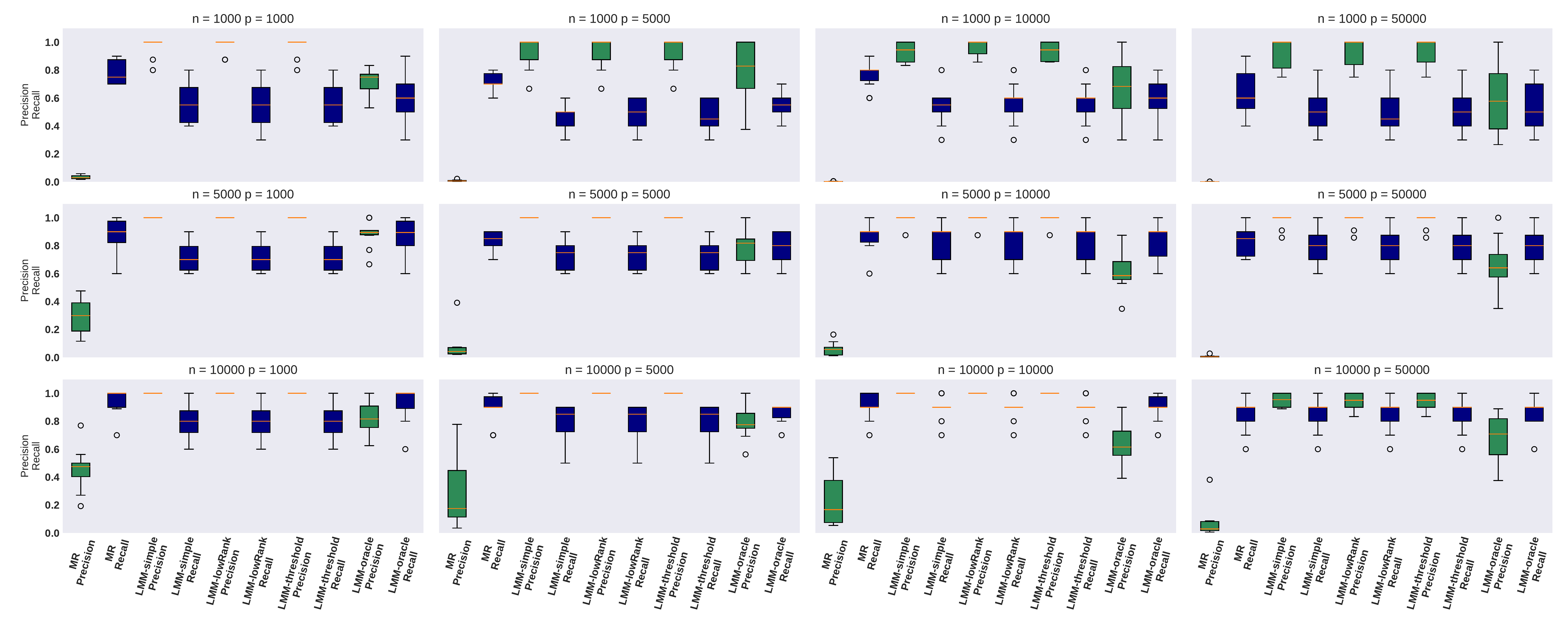}
    \caption{When population stratification are introduced, experiments are performed by altering $n$ and $p$. }
    \label{fig:exp4}
\end{figure}

\subsection{Approximating the Variance Component Introduced by Population Stratification}
\label{sec:exp:pop}

\quad 

\textbf{Data Generation:} 
We continue with the data generation process in \cref{sec:exp:exc} to introduce population stratification into the data. 
The SNP array $\mathbf{X}$ is generated in the same way as in Section~\ref{sec:exp:exc}, 
but since we do not compare to \lmf{} in this scenario, 
we enlarge the datasets to study with
$n=5000$, the data is simulated after 100 generations of mating of 5 populations, and the individuals of each population start to fall into 5 subpopulations after 80 generations. The random migrating rate is still 0.01. 
We then generate an unobserved SNP sequence (i.e. $\mathbf{U}$) of $k\cdot s$ associated SNPs, where $k$ is the number of associated SNPs in $\mathbf{X}$ and $s$ is a scaling factor. The phenotype is generated as:
$\mathbf{y} = \mathbf{X}\mathbf{\beta} + \mathbf{U}\mathbf{\alpha} + \mathbf{\epsilon}$
where $\mathbf{\alpha}$ is the corresponding effect sizes sampled following the same manner as $\mathbf{\beta}$. We fix $k=100$ and experiment with different settings such as: $s=\{1, 5, 10\}$ with $p = \{1000, 5000, 10000, 50000\}$. We sample the data with 10 different random seeds. 

\textbf{Association Study:} 
Again, Figure~\ref{fig:exp4} shows both the mean precision and mean recall of identifying the association SNP over 10 random seeds across the experiment settings.

\emph{LMMs in different situations:} As suggested by Proposition~\ref{lemma:k:pops}, whether LMM can help correct the confounding factors introduced through population stratification mainly depends on the configurations $n$ and $p$ of the SNP array. Therefore, we mainly study the different performances of \lms{} for different $n$ and $p$. We will also use the performances of \mr{} as a baseline for the expected performance drop when the problem gets harder as $p$ increases. Also, one should notice that Figure~\ref{fig:exp4} consists of different situations of low-dimension cases and high-dimension cases, and these situations need to be studied separately. 
With these clarifications, we can observe that (separate observations for low-dimension cases and high-dimension cases) as $p$ increases, the problem gets harder (performances of \mr{} drop), but the performances of \lms{} roughly maintains the same. 
We believe this performance preservation is due to the ability of \lms{} in correcting the confounding factors gets increased as $p$ increases. 
Also, we notice that the performances of \lms{}, \lml{}, and \lmt{}
are roughly the same for different situations in this section, 
which again validates our argument that 
the main ability of LMMs to correct population stratification depends on the configurations $n$ and $p$ of the SNP array. 

\begin{figure}
    \centering
    \includegraphics[width=1.0\textwidth]{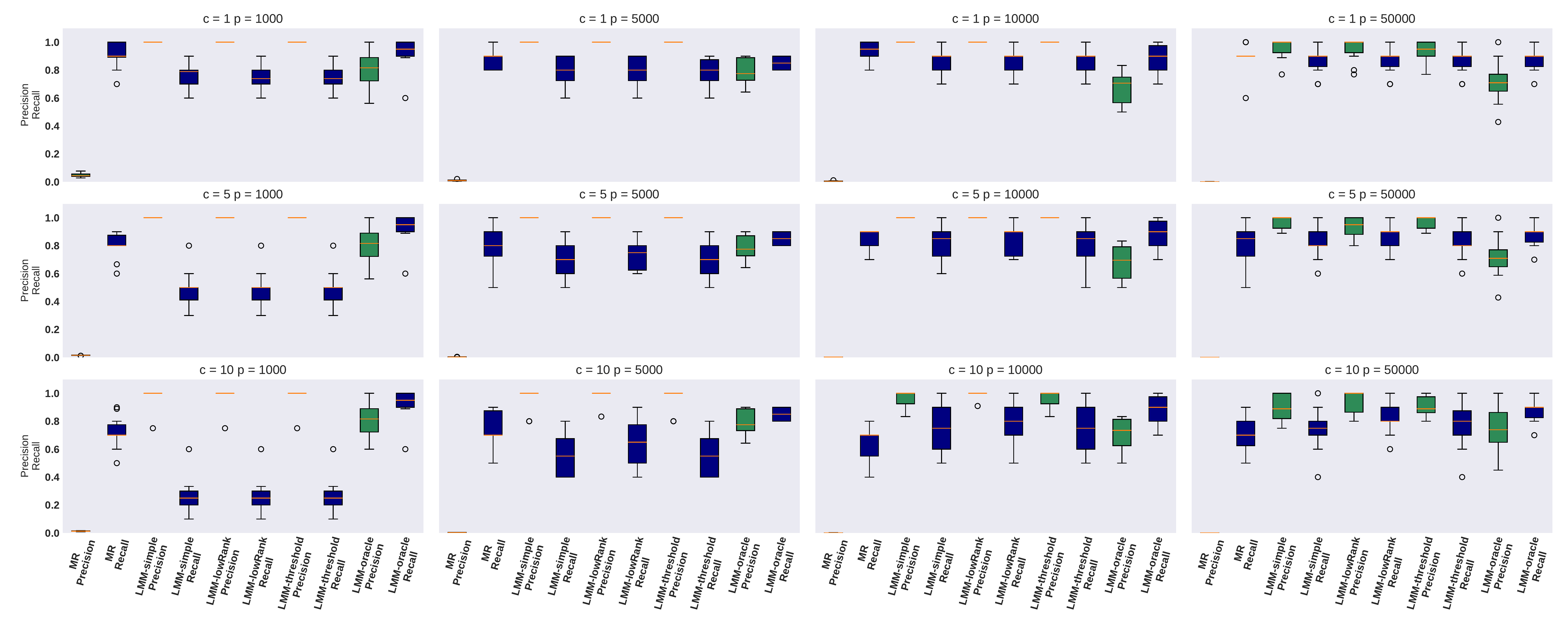}
    \caption{When environmental confounding factors on introduced, where $c$ denotes magnitudes of the effect sizes of the environmental confounding factor.}
    \label{fig:exp3}
\end{figure}

\subsection{Approximating the Variance Component Introduced by Environmental Confounding Factors}
\label{sec:exp:env}
Now, we study the effect of confounding factors introduced through the environment as empirical evidence to support the discussions in Proposition~\ref{theorem:K:main}.

\textbf{Data Generation:} 
We introduce the effects of confounding factors by
$\mathbf{y} = \mathbf{X}\mathbf{\beta} + \mathbf{D}\mathbf{\eta} + \mathbf{\epsilon}$
where $\mathbf{D}$ is the identification of the population structure. 
Specifically, each row of $\mathbf{D}$ is a one-hot vector whose elements are all zeros except one used to indicate the sample belongs to the corresponding population, and $\mathbf{\eta} \sim N(0, c^2\mathbf{I})$, where $c$ is a hyperparameter that controls the effects of the confounding factors. 
We fixed $k=100$, $n=500$, and $g=100$, and generated the simulation data with the settings $c\in\{1, 5, 10\}$ with $p \in \{1000, 5000, 10000, 50000\}$. 
Other details for generating $\mathbf{X}$ are the same as those in \cref{sec:exp:pop}. 
We sampled the data with 10 different random seeds for each case. 

\textbf{Association Study:} 
Similarly, Figure~\ref{fig:exp3} shows both the mean precision and mean recall of identifying the association SNP over 10 random seeds across the experiment settings. 

\emph{\lms{} vs. \lml{}:} As suggested by Proposition~\ref{theorem:K:main}, \lml{} generally reports a better performance, especially a better recall. 
Also, the advantage of \lml{} is barely observable when $c=1$, i.e. the effect of the environment factor is small but is more pronounced as $c$ gets larger. Therefore, it is likely that the performance gain of \lml{} in comparison to \lms{} is mostly due to its ability to correct confounding signals introduced by the environment. 


\emph{\lms{} vs. \lmt{}:} 
Using a fixed threshold of $\tau=0.001$ (the best value in terms of precision and recall we noticed across different trials), the advantages of \lmt{} over \lms{} seems marginal in most cases. However, this does not contradict Proposition~\ref{theorem:K:main} because we do not know the oracle $\tau$. 
To have a better understanding of \lmt{}, we also compare \lms{} and \lmo{}. 
In the low-dimension case, \lmo{} improves the recall significantly over \lms{}, while decreases precision reasonably as a trade-off. 
Overall, the improvement of \lmo{} over \lms{} and \lmt{} suggests that proper thresholding of the kinship matrix can indeed improve the selection of relevant SNPs, confirming Proposition~\ref{theorem:K:main}. 


%% file: secs/discussion.tex
Before we conclude, despite that our focus is the statistical aspects of LMM when applied to GWAS, 
we would like to devote several pragraph to discuss the practical aspects of the challenge of applying LMM in GWAS. 

\subsubsection{LMM and Multivariate Regression Methods}
Our results highlight the following crucial fact regarding mixed model methods applied to GWAS: While they are effective in controlling false positives in comparison with marginal regression, this improvement is also due to the well-known fact that LMM is essentially equivalent to Bayesian multiple regression (i.e. with Gaussian prior) that controls for the effects introduced by other SNPs as covariates \citep{maldonado2009mixed,heckerman2018accounting}.

Several interesting results regarding LMM in GWAS can be explained by this connection. For example, there is a branch of techniques that help to improve the empirical performance of LMM by using a
selective subset of SNP that are strongly associated with the trait \citep{listgarten2012improved,listgarten2013fast,wang2014super}. For example, \citet{listgarten2012improved} first used univariate method to select the SNPs that are strongly associated with trait and then used these SNPs to construct the variance component for LMM, leading to better empirical performance (in terms of controlling false positives) than the vanilla LMM set-up introduced in this paper. 
The connection between LMM and multivariate regression indicates that the empirical improvement may be because these techniques help identify a more accurate set of covariates to regress out from the trait, leading to a more accurate estimation of the effect size of the candidate SNP. 

Also, this connection helps address why LMMs seem to  consistently outperform marginal regression as a population stratification correction tool---even when population stratification may not always exist. This controversial point has been the subject of substantial debate in the genetics literature and is easily answered by statistical considerations alone \citep{wacholder2002counterpoint,freedman2004assessing,wang2004evaluating}. In fact, this paper was originally motivated by the observation that LMM can outperform marginal regression in terms of controlling false positives even if there are no explicit confounding factors in the data. These questions can be well answered by the by the fact that LMM behave similarly to multivariate methods. 

Further, we can also relate the connection to the widely usage of LMM with the construction of GRM with SNPs less correlated with the SNP of interest (\textit{i.e.}, to avoid proximal contamination). One of the most popular methods is probably to construct the GRM with SNPs that are not in the same chromosome with the SNP of interest, which is usually referred to as Leave-one-chromosome out (LOCO) \citep{yang2014advantages,listgarten2012improved}. 

Last but not least, if LMM behaves similarly to a multivariate regression method, then a natural follow-up question will be whether a multivariate regression method can replace LMM in GWAS. According to \citet{karkkainen2012robustness}, there is cumulative evidence that multivariate regression methods can capture the genetic relationships between individuals to account for confounding factors such as population stratification without explicitly building the variance component. 

\subsubsection{Case-control Studies}
Our discussion has been limited to quantitative (i.e. continuous) traits since this is the most common model used in the genetics literature. 
Directly applying the discussed LMM to case-control studies can potentially underestimate the true heritability \citep{golan2014measuring}. 
Fortunately, when dealing with case-control data, most GWAS studies consider the residual phenotypes, i.e., the residuals after the effect of covariates such as gender or age have been regressed out.
This practice is often used in GWAS studies \citep[e.g.,][]{atwell2010genome,valdar2006genome,liu2019association,wuttke2019catalog}. 
Further, although generalized linear mixed models (LMM for binary phenotype) have been extensively studied in the statistics literature, the complexity of the EM algorithm required to solve a generalized LMM seems to be a bottleneck of applying generalized LMM to GWAS. 
However, 
some specially designed LMMs have been invented for case-control data \citep[e.g.,][]{hayeck2015mixed,chen2016control}. 

\subsubsection{Other Modern LMMs for Specific Purposes}
Although the scope of this paper is limited to simple (i.e. vanilla) mixed models when applied to GWAS, many more advanced LMMs have been introduced. 
For example, FarmCPU \citep{liu2016iterative} iterates through estimation of fixed effect and random effect to improve estimation. 
Grid-LMM \citep{runcie2019fast} is introduced to fit multiple sources of variance components. 
\textsf{lme4qtl} \citep{ziyatdinov2018lme4qtl} introduces the flexibility of building custom variance components for LMM. 
REMMA \citep{ning2018rapid} and MAPIT \citep{crawford2017detecting} extends LMM for estimating epistasis and marginal epistasis respectively. 
There are also numerous methods incorporating various types of regularization such as the Lasso and other structured regularizers \citep{rakitsch2012lasso,Wang2019graph,Wang2019coupled}.


%% file: secs/appendix.tex
\section{Proof of Theorem~3.1}

The proof of Theorem 3.1 mainly involves the following two lemmas: 

\begin{lemma}
For all $s>0$, 
$\widehat{\beta}(\mathbf{K}; s) = \widehat{\beta}(\mathbf{K'}; s)$
where $\mathbf{K}=\dfrac{1}{n}\mathbf{Z}\mathbf{Z}^T$ and $\mathbf{K}' = \dfrac{1}{n}[\mathbf{Z}, \mathbf{x}][\mathbf{Z}, \mathbf{x}]^T$
\label{lemma:rr:eq}
\end{lemma}
\begin{proof}
With $\mathbf{K}=\dfrac{1}{n}\mathbf{Z}\mathbf{Z}^T$, and $\mathbf{K}' = \dfrac{1}{n}[\mathbf{Z}, \mathbf{x}][\mathbf{Z}, \mathbf{x}]^T = \dfrac{1}{n}\mathbf{Z}\mathbf{Z}^T + \dfrac{1}{n}\mathbf{x}\mathbf{x}^T$, 
with Sherman-Morrison Formula, 
\begin{align*}
\mathbf{M}' & = (\mathbf{M}^{-1} + \dfrac{1}{n}\mathbf{x}\mathbf{x}^T)^{-1} \\
&= n(n\mathbf{M}^{-1}+\mathbf{x}\mathbf{x}^T)^{-1} \\
&= n(\dfrac{1}{n}\mathbf{M} - \dfrac{\frac{1}{n}\mathbf{M}\mathbf{x}\mathbf{x}^T\mathbf{M}}{1+\frac{1}{n}\mathbf{x}^T\mathbf{M}\mathbf{x}})\\
& = \mathbf{M} - n\dfrac{\mathbf{M}\mathbf{x}\mathbf{x}^T\mathbf{M}}{n + \mathbf{x}^T\mathbf{M}\mathbf{x}}
\end{align*}
where $\mathbf{M}=(s \mathbf{I} + \dfrac{1}{n}\mathbf{Z}\mathbf{Z}^T)^{-1}$ and $\mathbf{M}'=(s \mathbf{I} + \dfrac{1}{n}\mathbf{Z}\mathbf{Z}^T + \dfrac{1}{n}\mathbf{x}\mathbf{x}^T)^{-1}$. 

Now we have:
\begin{align*}
    \widehat{\beta}(\mathbf{K'}; s) &= \dfrac{\mathbf{x}^T\mathbf{M}'\mathbf{y}}{\mathbf{x}^T\mathbf{M}'\mathbf{x}} \\
    & = \dfrac{\mathbf{x}^T\mathbf{M}\mathbf{y}-\frac{n}{n + \mathbf{x}^T\mathbf{M}\mathbf{x}}(\mathbf{x}^T\mathbf{M}\mathbf{x}\mathbf{x}^T\mathbf{M}\mathbf{y})}{\mathbf{x}^T\mathbf{M}\mathbf{x}-\frac{n}{n + \mathbf{x}^T\mathbf{M}\mathbf{x}}(\mathbf{x}^T\mathbf{M}\mathbf{x}\mathbf{x}^T\mathbf{M}\mathbf{x})} \\
    & = \dfrac{\mathbf{x}^T\mathbf{M}\mathbf{y}(\frac{n + (1 - n)\mathbf{x}^T\mathbf{M}\mathbf{x}}{n + \mathbf{x}^T\mathbf{M}\mathbf{x}})}{\mathbf{x}^T\mathbf{M}\mathbf{x}(\frac{n + (1 - n)\mathbf{x}^T\mathbf{M}\mathbf{x}}{n + \mathbf{x}^T\mathbf{M}\mathbf{x}})}\\
    & = \dfrac{\mathbf{x}^T\mathbf{M}\mathbf{y}}{\mathbf{x}^T\mathbf{M}\mathbf{x}} = \widehat{\beta}(\mathbf{K}; s)
\end{align*}
\end{proof}

\begin{lemma}
\label{lemma:rr:2}
For any $\Delta \geq -s$, 
\begin{align*}
\dfrac{1 + \svmin(s)\Delta}{1 + \svmax(s)\Delta}
\leq \dfrac{\est(\bZ\bZ^{T}; s+\Delta)}{\est(\bZ\bZ^{T}; s)} 
\leq
\dfrac{1 + \svmax(s)\Delta}{1 + \svmin(s)\Delta}, 
\end{align*}
\end{lemma}

\begin{proof}
We only consider the case $\est(\bZ\bZ^{T}; s)$, the other case can be derived following the same procedure. 

We start with $\est(\bZ\bZ^{T}; s) = \dfrac{\mathbf{x}^T\mathbf{M}\mathbf{y}}{\mathbf{x}^T\mathbf{M}\mathbf{x}}$ where $\mathbf{M}=(s \mathbf{I} + \dfrac{1}{n}\mathbf{Z}\mathbf{Z}^T)^{-1}$, and $\mathbf{M}'=(\mathbf{M}^{-1}+\Delta \mathbf{I})^{-1}$, so we have:
\begin{align*}
\est(\bZ\bZ^{T}; s + \Delta)= \dfrac{\mathbf{x}^T\mathbf{M}'\mathbf{y}}{\mathbf{x}^T\mathbf{M}'\mathbf{x}}
\end{align*}

Due to the construction of $\mathbf{M}$, we conveniently have the following equations hold through singular value decomposition:
\begin{align*}
\mathbf{M} &= \mathbf{U}\mathbf{S}\mathbf{V}^T, \quad \textnormal{where} \quad  \mathbf{V}=\mathbf{U} \\ 
\mathbf{M}' &=\mathbf{U}\mathbf{S}'\mathbf{V}^T, \quad \textnormal{where} \quad \mathbf{S}' = (\mathbf{S}^{-1} + \Delta\mathbf{I})^{-1} 
\end{align*}
and $\mathbf{S}_{\textnormal{max}} \leq 1/\delta $. 

Further, we introduce the notations of $\mathbf{r}$ and $\mathbf{t}$ for convenience, where $\mathbf{r}=\mathbf{V}^Tx$ and $\mathbf{t}=\mathbf{V}^Ty$.
Therefore, we have:
\begin{align*}
\est(\bZ\bZ^{T}; s + \Delta) =& \dfrac{\mathbf{x}^T\mathbf{M}'\mathbf{y}}{\mathbf{x}^T\mathbf{M}'\mathbf{x}} = \dfrac{\sum_i\mathbf{r}_i^T\mathbf{S}_i'\mathbf{t}}{\sum_i\mathbf{r}_i^T\mathbf{S}_i'\mathbf{r}} \\
=& \dfrac{\sum_i\mathbf{r}_i^T\dfrac{1}{1/\mathbf{S}_i+\Delta }\mathbf{t}_i}{\sum_i\mathbf{r}_i^T\dfrac{1}{1/\mathbf{S}_i+\Delta}\mathbf{r}_i} \\
=& \dfrac{\sum_i\dfrac{\mathbf{r}_i^T\mathbf{S}_i\mathbf{t}_i}{1+\Delta \mathbf{S}_i}}{\sum_i\dfrac{\mathbf{r}_i^T\mathbf{S}_i\mathbf{r}_i}{1+\Delta \mathbf{S}_i}} 
\end{align*}
We extract the term $1 + \Delta \mathbf{S}_i$ out from both numerator and denominator, and we have:
\begin{align*}
    \dfrac{1+\Delta \mathbf{S}_i |_{\min}}{1+\Delta \mathbf{S}_i |_{\max}} \est(\bZ\bZ^{T}; s) \leq \est(\bZ\bZ^{T}; s + \Delta) 
    \leq \dfrac{1+\Delta \mathbf{S}_i |_{\max}}{1+\Delta \mathbf{S}_i |_{\min}}\est(\bZ\bZ^{T}; s),
\end{align*}
where $1+\Delta \mathbf{S}_i |_{\min}$ (or $1+\Delta \mathbf{S}_i |_{\max}$) denotes the minimum (or maximum) value of $1+\Delta \mathbf{S}_i$. Due to the construction of $\mathbf{M}$, we have $0 < \mathbf{S}_i \leq 1/s$, so for any $\Delta > -s$, we have $1+\Delta \mathbf{S}_i > 0$ for any $i$. 
In other words, we have $1+\Delta \mathbf{S}_i > 0$ monotonically increasing as a function of $\mathbf{S}_i$ for any $\Delta > -s$.  

Therefore, for any $\Delta > -s$, we have:
\begin{align*}
    \dfrac{1+\Delta \svmin(s)}{1+\Delta \svmax(s)}\est(\bZ\bZ^{T}; s) \leq \est(\bZ\bZ^{T}; s + \Delta) \leq \dfrac{1+\Delta \svmax(s)}{1+\Delta \svmin(s)}\est(\bZ\bZ^{T}; s)
\end{align*}
\end{proof}

Combining the two Lemmas above will lead us to (8). We sketch the argument here. In Theorem 1, we need to require $\Delta > -\widehat{\delta}(\mathbf{Z}\mathbf{Z}^T)$ to align with the condition $\Delta >-s$ in Lemma S1.2. 
Consider the case $\est(\bZ\bZ^{T}) > 0$. 
After rearranging the terms, we have:
\begin{align}
    \dfrac{\est(\bZ\bZ^{T} + \bx\bx^{T})}{\est(\bZ\bZ^{T})} \leq \dfrac{1 + \Delta \svmax}{1 + \Delta \svmin}.
\end{align}
Now consider when 
\begin{align}
\label{eq:2}
    \dfrac{1 + \Delta \svmax}{1 + \Delta \svmin} < 1 + \epsilon
\end{align}
for some $\epsilon>0$.

With some algebra, assuming that $\svmax - (1+\epsilon)\svmin> 0$, we have: 
\begin{align}
\label{eq:delta:bound:first}
    \Delta  < \dfrac{\epsilon}{\svmax - (1+\epsilon)\svmin}
    :=RHS,
\end{align}
Also, since RHS $\ge \epsilon/(\svmax-\svmin)$, it suffices to study 
\begin{align*}
    \Delta \leq \dfrac{\epsilon}{\svmax-\svmin} 
\end{align*}
By definition, we have: 
\begin{align*}
    \svmax &= \dfrac{1}{\widehat{\delta}_{[\mathbf{Z}]} + \mathbf{A}_{\textnormal{min}}} \\
    \svmin &= \dfrac{1}{\widehat{\delta}_{[\mathbf{Z}]} + \mathbf{A}_{\textnormal{max}}}
\end{align*}
where we use $\mathbf{A}$ to denote the singular values of $\dfrac{1}{n}\mathbf{Z}^T\mathbf{Z}$. Thus, now we have: 
\begin{align*}
    \Delta < \dfrac{(\mathbf{A}_{\textnormal{max}} + \widehat{\delta}(\bZ\bZ^T))(\mathbf{A}_{\textnormal{min}} + \widehat{\delta}(\bZ\bZ^T))}{\mathbf{A}_{\textnormal{max}} - \mathbf{A}_{\textnormal{min}}}\epsilon,
\end{align*}
which means $\Delta $ needs to be upper bounded by 
\begin{align*}
    \dfrac{(\mathbf{A}_{\textnormal{max}} + \widehat{\delta}(\bZ\bZ^T))(\mathbf{A}_{\textnormal{min}} + \widehat{\delta}(\bZ\bZ^T))}{\mathbf{A}_{\textnormal{max}} - \mathbf{A}_{\textnormal{min}}}\epsilon
\end{align*}
for $\est(\bZ\bZ^{T} + \bx\bx^{T})/\est(\bZ\bZ^{T})$ to be bounded by $1+\epsilon$. 
This is precisely $\htol(\bZ)$. The lower bound is proved similarly.

Now we continue to prove a similar result for the standardized coefficient estimates: 
\begin{align}
\label{eq:rr:main3}
\dfrac{1}{(1+\epsilon)^2}
< \dfrac{p(\bZ\bZ^{T} + \bx\bx^{T})}{p(\bZ\bZ^{T})} 
< (1+\epsilon)^2,
\end{align}
where $p(\mathbf{K}) = \est^2(\mathbf{K})/\Var(\est(\mathbf{K}))$.
The proof is similar to the proof of (8); we sketch out the details below.
\begin{lemma}
For all $s>0$, 
$\Var(\widehat{\beta}(\mathbf{K}; s)) = \Var(\widehat{\beta}(\mathbf{K'}; s))$
where $\mathbf{K}=\dfrac{1}{n}\mathbf{Z}\mathbf{Z}^T$ and $\mathbf{K}' = \dfrac{1}{n}[\mathbf{Z}, \mathbf{x}][\mathbf{Z}, \mathbf{x}]^T$
\label{lemma:rr:eq:standardized}
\end{lemma}
\begin{proof}
The proof follows the same procedure as the one in Lemma S1.1 by noting that 
\begin{align*}
    \Var(\widehat{\beta}(\mathbf{K}; s)) &= \mathbb{E}(\widehat{\beta}^2(\mathbf{K}; s)) - \mathbb{E}^2(\widehat{\beta}(\mathbf{K}; s)) \\
    & = \mathbb{E}((\dfrac{\mathbf{x}^T\mathbf{M}\mathbf{y}}{\mathbf{x}^T\mathbf{M}\mathbf{x}})^2) - \mathbb{E}^2(\dfrac{\mathbf{x}^T\mathbf{M}\mathbf{y}}{\mathbf{x}^T\mathbf{M}\mathbf{x}}) \\
    & = \mathbb{E}((\dfrac{\mathbf{x}^T\mathbf{M}(\mathbf{x}\beta + \randeff + \epsilon)}{\mathbf{x}^T\mathbf{M}\mathbf{x}})^2) - \mathbb{E}^2(\dfrac{\mathbf{x}^T\mathbf{M}(\mathbf{x}\beta + \randeff + \epsilon)}{\mathbf{x}^T\mathbf{M}\mathbf{x}})\\
    &= \dfrac{\mathbf{x}^T\mathbf{M}^2\mathbf{x}}{(\mathbf{x}^T\mathbf{M}\mathbf{x})^2}\sigma_\mathbf{y}
\end{align*}
$\randeff$ and $\epsilon$ denote the random effects and residual noises respectively, as introduced in (1) in the main paper. 
We use $\sigma_\mathbf{y}$ to denote $\textnormal{Var}(\randeff + \epsilon)$. 

The rest of the proof follows the same procedure as in the proof for Lemma 1.1.
\end{proof}

\begin{lemma}
\label{lemma:rr:2:standardized}
For any $\Delta \geq -s$, 
\begin{align*}
(\dfrac{1 + \svmin(s)\Delta}{1 + \svmax(s)\Delta})^2
\leq \dfrac{p(\bZ\bZ^{T}; s+\Delta)}{p(\bZ\bZ^{T}; s)}
\leq
(\dfrac{1 + \svmax(s)\Delta}{1 + \svmin(s)\Delta})^2.
\end{align*}
\end{lemma}
\begin{proof}
Following the definition, we have
\begin{align*}
    p(\bZ\bZ^{T}; s) = \dfrac{\est^2(\bZ\bZ^{T}; s)}{\Var(\est(\bZ\bZ^{T}; s))} = \dfrac{(\mathbf{x}^T\mathbf{M}\mathbf{y})^2}{\mathbf{x}^T\mathbf{M}^2\mathbf{x}\sigma_\mathbf{y}}.
\end{align*}
The rest of the proof follows the same procedure as the one in Lemma S1.2, 
which can be sketched as:
We first rewrite $\mathbf{M}'$ with $\mathbf{U}$, $\mathbf{V}$, and $\mathbf{S}$, then expand $p(\bZ\bZ^{T}; s+\Delta)$ with $\mathbf{r}$, $\mathbf{t}$, and $\mathbf{S}$, and we have
\begin{align*}
    p(\bZ\bZ^{T}; s+\Delta) = \dfrac{\sum_i\dfrac{\mathbf{r}_i^T\mathbf{S}_i\mathbf{t}_i}{1+\Delta \mathbf{S}_i}\sum_i\dfrac{\mathbf{r}_i^T\mathbf{S}_i\mathbf{r}_i}{1+\Delta \mathbf{S}_i}}{\sum_i\dfrac{\mathbf{r}_i^T\mathbf{S}_i^2\mathbf{r}_i}{(1+\Delta \mathbf{S}_i)^2}\sigma_\mathbf{y}}
\end{align*}
Instead of extracting the term $1 + \Delta\mathbf{S}_i$, we extract $(1 + \Delta\mathbf{S}_i)^2$, 
and then, the remaining arguments follow the ones in the proof of Lemma S1.2. 
\end{proof}
Combining the two lemmas above yields
\begin{align*}
   (\dfrac{1 + \svmin(s)\Delta}{1 + \svmax(s)\Delta})^2 \leq \dfrac{p(\bZ\bZ^{T} + \bx\bx^{T})}{p(\bZ\bZ^{T})} \leq (\dfrac{1 + \svmax(s)\Delta}{1 + \svmin(s)\Delta})^2.
\end{align*}
From here, it suffices to deduce when \eqref{eq:2} holds, which follows the same argument as before.

\section{Proof of Theorem~3.2}

\begin{proof}
Follow Theorem 2.16 in \citep{bai2008methodologies}, which states the almost surly convergence of extreme eigenvalues, together with the facts that 1) positive semidefinite symmetric matrix's eigenvalues coincide the singular values, and 2) almost sure convergence implies convergence in probability by Fatou's lemma, we have:
\begin{align*}
    \mathbf{A}_{\textnormal{max}} \xrightarrow{p} (1 + \sqrt{\zeta})^2 \\
    \mathbf{A}_{\textnormal{min}} \xrightarrow{p} (1 - \sqrt{\zeta})^2
\end{align*}
where $\dfrac{q+1}{n} \xrightarrow{} \zeta $.  

Then, with continuous mapping theorem, we have:
\begin{align*}
    \Gamma & \xrightarrow{p} \dfrac{((1 + \sqrt{\zeta})^2 + \widehat{\delta}_{[\mathbf{Z}, \bx]}) ((1 - \sqrt{\zeta})^2 + \widehat{\delta}_{[\mathbf{Z}, \bx]})}{(1 + \sqrt{\zeta})^2 - (1 - \sqrt{\zeta})^2} \\
    & = \dfrac{(\zeta + \widehat{\delta}_{[\mathbf{Z}, \bx]})^2 + 2\widehat{\delta}_{[\mathbf{Z}, \bx]} - 2\zeta + 1}{4\sqrt{\zeta}}
\end{align*}

According to \citep{jiang2016high}, $\widehat{\delta}_{[\mathbf{Z}, \bx]} \xrightarrow{p} \dfrac{1}{\omega}\delta_{[\mathbf{Z}, \bx]}$, where $\omega = \lim_{n\rightarrow \infty }\dfrac{k}{q+1}$ and $k$ denotes the number of associated SNPs, therefore, with continuous mapping theorem, we can have:
\begin{align*}
    \Gamma \xrightarrow{p} \dfrac{(\zeta + \dfrac{\delta_{[\mathbf{Z}]}}{\omega})^2 + \dfrac{2\delta_{[\mathbf{Z}, \bx]}}{\omega} - 2\zeta + 1}{4\sqrt{\zeta}},
\end{align*}
\end{proof}

\section{Proof of Proposition 3.3}
\begin{proof}
We first consider that $\mathbf{K} =\sum_t \dfrac{1}{q}\mathbf{Z}_t\mathbf{Z}_t^T$ where $t$ is the index of SNPs. By definition, we have:
\begin{align*}
  \Vert\dfrac{1}{q}\mathbf{Z}_t\mathbf{Z}_t^T -\mathbb{E}[\dfrac{1}{q}\mathbf{Z}_t\mathbf{Z}_t^T]\Vert_2 \leq \Vert\dfrac{1}{q}\mathbf{Z}_t\mathbf{Z}_t^T\Vert_2 \leq \dfrac{4n}{q}
\end{align*}

Now we consider $\mathbf{J}=\mathbf{K}-\mathbb{E}[\mathbf{K}]=\mathbf{K}-\mathbb{E}[\truecov_{\textnormal{pop}}]$ (because of $\mathbb{E}[\mathbf{K}] = \mathbb{E}[\truecov_{\textnormal{pop}}]$ from the assumption), and $\mathbb{E}[\mathbf{J}] = \mathbf{0}$ by definition. We decompose $\mathbf{J}$ as the sum of $q$ matrices 
\begin{align*}
    \mathbf{J} =\sum_t\mathbf{J}_t = \sum_t \dfrac{1}{q}\mathbf{Z}_t\mathbf{Z}_t^T -\mathbb{E}[\truecov_{\textnormal{pop}}]
\end{align*}
And we have the bound $\Vert\mathbf{J}_t\Vert_2 \leq \dfrac{4n}{q}$

In addition, by the definition of the SNP data, we can have:
\begin{align*}
    \mathbb{E}[\Vert\mathbf{J}_t^T\mathbf{J}_t\Vert_2] = \mathbb{E}[\Vert\mathbf{J}_t\mathbf{J}_t^T\Vert_2] \leq \dfrac{16n^2}{q^2}
\end{align*}
thus,
\begin{align*}
    \Vert\sum_t\mathbb{E}[\mathbf{J}_t^T\mathbf{J}_t]\Vert_2 \leq \sum_t\Vert\mathbb{E}[\mathbf{J}_t^T\mathbf{J}_t]\Vert_2 = \dfrac{16n^2}{q}
\end{align*}

By applying Matrix Bernstein Theorem \citep{tropp2015introduction}, we can have:
\begin{align*}
    \mathbb{P}(\Vert\mathbf{K} - \mathbb{E}[\truecov_{\textnormal{pop}}]\Vert_2 \geq t) & \leq 2n\exp(-\dfrac{t^2/2}{16n^2/q + 4nt/3q}) \\
    & = 2n\exp(-\dfrac{t^2}{32n^2/q + 8nt/3q}) \\ 
    & = 2n\exp(-\dfrac{3t^2q}{96n^2 + 8nt})
\end{align*}
\end{proof}

\section{Proof of Proposition 3.4}
\begin{proof}
In order to study the approximation error, we need to study the differences between $\mathbf{K}$ and $\mathbf{G}_{\textnormal{env}}(\rho)$. 

The result can be simply proved if we write out the expressions. 
With the definitions, we can directly write 
\begin{align*}
    \Vert\mathbb{E}[\mathbf{K}] - \mathbf{G}_{\textnormal{env}}(\rho)\Vert_F^2 &= \sum_{i,j|i,j\in \mathcal{C}}(\mathbb{E}[\mathbf{K}_{i,j}] - \rho)^2 + \sum_{i,j|i,j\not\in \mathcal{C}} (\mathbb{E}[\mathbf{K}_{i,j}])^2 \\
    \Vert\mathbb{E}[\tilde{\mathbf{K}}] - \mathbf{G}_{\textnormal{env}}(\rho)\Vert_F^2 &= \sum_{i,j|i,j\in \mathcal{C}}(\mathbb{E}[\tilde{\mathbf{K}}_{i,j}] - \rho)^2 + \sum_{i,j|i,j\not\in \mathcal{C}} (\mathbb{E}[\tilde{\mathbf{K}}_{i,j}])^2 \\
    \Vert\mathbb{E}[\grave{\mathbf{K}}(\tau^\star)] - \mathbf{G}_{\textnormal{env}}(\rho)\Vert_F^2 &= \sum_{i,j|i,j\in \mathcal{C}}(\mathbb{E}[\grave{\mathbf{K}}(\tau^\star)_{i,j}] - \rho)^2 + \sum_{i,j|i,j\not\in \mathcal{C}} (\mathbb{E}[\grave{\mathbf{K}}(\tau^\star)_{i,j}])^2
\end{align*}
With oracle $\tau^\star$, we have 
\begin{align*}
    \Vert\mathbb{E}[\grave{\mathbf{K}}(\tau^\star)] - \mathbf{G}_{\textnormal{env}}(\rho)\Vert_F^2 &= \sum_{i,j|i,j\in \mathcal{C}}(\mathbb{E}[\mathbf{K}_{i,j}] - \rho)^2 
\end{align*}
With Assumptions 8 in the main paper, 
for the first inequality of Assumption 8, 
we have:
\begin{align*}
    (\mathbb{E}[\tilde{\mathbf{K}}_{i,j|(i,j) \not\in \mathcal{C}}])^2 - (\mathbb{E}[\mathbf{K}_{i,j|(i,j) \not\in \mathcal{C}}])^2
    \leq 
    (\mathbb{E}[\mathbf{K}_{i,j|(i,j) \in \mathcal{C}}]-\rho)^2 - 
    (\mathbb{E}[\tilde{\mathbf{K}}_{i,j|(i,j) \in \mathcal{C}}]-\rho)^2 
\end{align*}
which is equivalently
\begin{align*}
    (\mathbb{E}[\tilde{\mathbf{K}}_{i,j|(i,j) \not\in \mathcal{C}}])^2 +
    (\mathbb{E}[\tilde{\mathbf{K}}_{i,j|(i,j) \in \mathcal{C}}]-\rho)^2
    \leq 
    (\mathbb{E}[\mathbf{K}_{i,j|(i,j) \not\in \mathcal{C}}])^2 + 
    (\mathbb{E}[\mathbf{K}_{i,j|(i,j) \in \mathcal{C}}]-\rho)^2
\end{align*}
which is equivalently
\begin{align*}
    \Vert\mathbb{E}[\tilde{\mathbf{K}}] - \mathbf{G}_{\textnormal{env}}(\rho)\Vert_F^2 \leq 
    \Vert\mathbb{E}[\mathbf{K}] - \mathbf{G}_{\textnormal{env}}(\rho)\Vert_F^2. 
\end{align*}
Similarly, for the second inequality of Assumption 8, we have
\begin{align*}
    (\mathbb{E}[\mathbf{K}_{i,j|(i,j) \in \mathcal{C}}]-\rho)^2 - 
    (\mathbb{E}[\tilde{\mathbf{K}}_{i,j|(i,j) \in \mathcal{C}}]-\rho)^2
    &\leq 
    (\mathbb{E}[\tilde{\mathbf{K}}_{i,j|(i,j) \not\in \mathcal{C}}])^2
\end{align*}
which is equivalently
\begin{align*}
    (\mathbb{E}[\mathbf{K}_{i,j|(i,j) \in \mathcal{C}}]-\rho)^2 
    &\leq 
    (\mathbb{E}[\tilde{\mathbf{K}}_{i,j|(i,j) \not\in \mathcal{C}}])^2 
    + 
    (\mathbb{E}[\tilde{\mathbf{K}}_{i,j|(i,j) \in \mathcal{C}}]-\rho)^2, 
\end{align*}
which is equivalently
\begin{align*}
    \Vert\mathbb{E}[\grave{\mathbf{K}}(\tau^\star)] - \mathbf{G}_{\textnormal{env}}(\rho)\Vert_F^2 \leq 
    \Vert\mathbb{E}[\tilde{\mathbf{K}}] - \mathbf{G}_{\textnormal{env}}(\rho)\Vert_F^2
\end{align*}
Therefore, we have
\begin{align*}
    \Vert\mathbb{E}[\grave{\mathbf{K}}(\tau^\star)] - \mathbf{G}_{\textnormal{env}}(\rho)\Vert_F^2 \leq 
    \Vert\mathbb{E}[\tilde{\mathbf{K}}] - \mathbf{G}_{\textnormal{env}}(\rho)\Vert_F^2 \leq 
    \Vert\mathbb{E}[\mathbf{K}] - \mathbf{G}_{\textnormal{env}}(\rho)\Vert_F^2
\end{align*}
\end{proof}

\newpage 

\begin{figure}[!htb]
\centering 
\minipage{0.4\textwidth}
  \includegraphics[width=\linewidth]{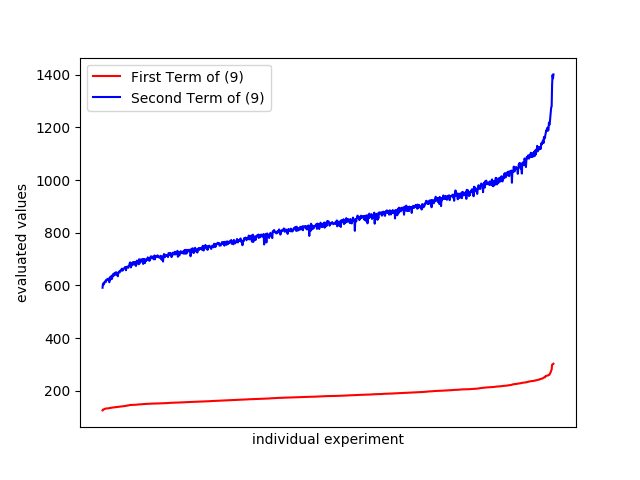}
\endminipage
\quad\quad 
\minipage{0.4\textwidth}
  \includegraphics[width=\linewidth]{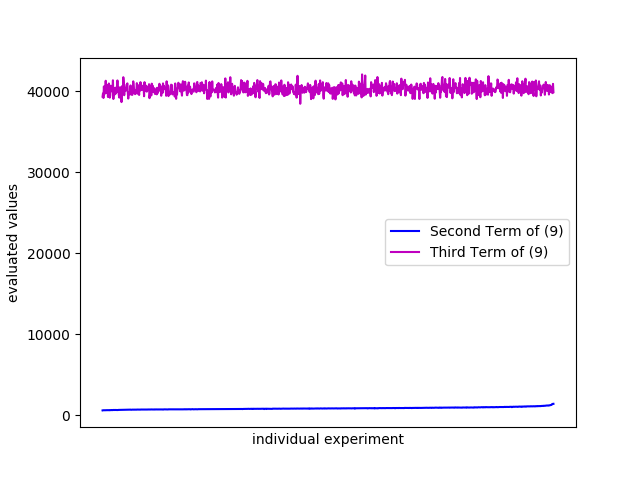}
\endminipage
\caption{Visualization of the empirical validation of Assumption 9}\label{fig:assumption}
\end{figure}

\section{Validation of Assumption 8 for Proposition 4}
Using simulated real-world data from SimuPop \citep{peng2005simupop}, we explicitly check this condition as follows: We used SimuPop to sample 1000 SNP arrays of 500 samples and 5000 SNPs with 5 generations of random matching.
With the simulated SNP array and assumed population structure $\mathbf{G}_{env}(\rho)$, we can explicitly calculate the three terms 
in Assumption 8. 
For the terms related to the $\tilde{\mathbf{K}}$, we can follow \citep{wang2017variable} to reconstruct the low-rank approximation matrix. 
The results of the 1000 experiments are shown in Figure~\ref{fig:assumption}.
(Figure~\ref{fig:assumption}(left) for comparison between the first term and the second term, and Figure~\ref{fig:assumption}(right) for comparison between the second term and the third term). 
The clear gap between the three terms provides evidence that Assumption 8 holds in real data with a strong population structure, indicating that low-rank approximations provide better approximations to $\mathbf{G}_{\textnormal{env}}(\rho)$ in practice. In other words, assuming the data does in fact have environmental confounding factors, then the vanilla LMM is suboptimal, and can be replaced with a low-rank approximation that will improve estimation. Note that although we have focused only on the approximation error between the kinship and $\mathbf{G}_{\textnormal{env}}(\rho)$, it is straightforward to include terms for the estimation error (i.e. $\mathbf{K}-\mathbb{E}[\mathbf{K}]$) and the approximation error between $\mathbf{G}_{\textnormal{env}}(\rho)$ and the \emph{true} population covariance.

Finally, we note also that the first inequality builds upon the general understanding of low-rank approximation methods:
Intuitively, the assumption suggests that the low-rank approximation will penalize the entries off the block-diagonal more than the entries within the block-diagonal structure. This is a standard observation, for example in the image denoising literature \citep{guo2015efficient}.